\documentclass[aps,pra,twocolumn,groupedaddress,eqsecnum]{revtex4-1}

\usepackage{graphicx} 
\usepackage{color} 
\usepackage{amsthm}
\usepackage{amsmath}
\usepackage{amssymb}
\usepackage{amsfonts}
\usepackage{setspace}
\usepackage{nomencl}
\usepackage{amstext}
\usepackage{comment}

\newcommand{\uphi}{\boldsymbol{\phi}}
\newcommand{\utheta}{\boldsymbol{\theta}}
\newcommand{\uthetat}{\widetilde{\boldsymbol{\theta}}}
\newcommand{\uthetahat}{\hat{\boldsymbol{\theta}}}

\begin{document}

\title{Investigating the limits of randomized benchmarking protocols}

\author{Jeffrey M. Epstein}
\affiliation{IBM T.J. Watson Research Center, Yorktown Heights, NY 10598, USA}
\author{Andrew W. Cross}
\affiliation{IBM T.J. Watson Research Center, Yorktown Heights, NY 10598, USA}
\author{Easwar Magesan}
\affiliation{IBM T.J. Watson Research Center, Yorktown Heights, NY 10598, USA}
\author{Jay M. Gambetta}
\affiliation{IBM T.J. Watson Research Center, Yorktown Heights, NY 10598, USA}


\begin{abstract}
In this paper, we analyze the performance of randomized benchmarking protocols on gate sets under a variety of realistic error models that include systematic rotations, amplitude damping, leakage to higher levels, and 1/f noise. We find that, in almost all cases, benchmarking provides better than a factor-of-two estimate of average error rate, suggesting that randomized benchmarking protocols are a valuable tool for verification and validation of quantum operations. In addition, we derive new models for fidelity decay curves under certain types of non-Markovian noise models such as 1/f and leakage errors. We also show that, provided the standard error of the fidelity measurements is small, only a small number of trials are required for high confidence estimation of gate errors.
\end{abstract}


\maketitle

\section{Introduction}

The advancement of experimental quantum information processing requires a method to benchmark errors on quantum gates. These benchmarks provide straightforward methods for comparing different experimental implementations and also establish compliance with error thresholds for processes such as error correction \cite{Cross2009}. The standard method for characterizing errors is quantum process tomography (QPT) \cite{Chuang1997,Poyatos1997}, which provides complete error reconstruction. Implementing QPT comes at a significant price though, since its complexity scales exponentially in order to determine the $16^n$ real parameters of the $n$-qubit quantum error process. In addition, QPT is vulnerable to state preparation and measurement (SPAM) errors, which are the errors associated with preparing and measuring different states. Because these errors may be on the same order as the error on the gate of interest, they can cause significant inaccuracies in the reconstructed errors. A recent study found that QPT overestimated small errors by several orders of magnitude \cite{Merkel2013}. 

An alternative to QPT is randomized benchmarking (RB) \cite{Knill2008,Magesan2011,Magesan2012,Magesan2012a,Gaebler2012,Gambetta2012,Corcoles2013}. Because this method extracts specific parameters of interest from the noise, as opposed to the complete set of parameters obtained from QPT, it does not suffer from exponential scaling. RB is also impervious to SPAM errors since it examines fidelity decays over random gate sequences. RB protocols have become an important tool for quantum verification and validation, and have been used to benchmark one- and two-qubit gates in atomic ion \cite{Knill2008,Biercuk2009,Brown2011,Olmschenk2010,Gaebler2012}, NMR \cite{Ryan2009}, and superconducting qubit \cite{Chow2009,Chow2010a,Magesan2012a,Gambetta2012,Corcoles2013} experiments. RB protocols and their fidelity decay models are provably valid in a wide variety of scenarios, however assumptions about the form of the fidelity decay under complex noise models may introduce inaccuracies. As well, there have been concerns about the convergence of the estimate from finite sampling effects since typical experiments are performed with many fewer random sequences then predicted using the Hoeffding bound~\cite{Magesan2011}. For these reasons, it is important to study and develop new models in the limits where analytic support for current RB methods may be lacking. 

In this paper, we address many of these issues by providing new theoretical results on finite sampling effects and modeling decay curves for realistic types of noise. As well, we test the performance of RB in different scenarios by implementing numerical simulations of both standard~\cite{Knill2008,Magesan2011} and interleaved~\cite{Magesan2012a,Gaebler2012} RB protocols under several physically relevant single-qubit error models. Two main classes of noise were tested; Markovian and non-Markovian. Markovian noise is memory-less and as such is history-independent, while non-Markovian noise is history-dependent, so the noise at one moment may depend on previous gates in the sequence. The first type of Markovian noise we investigated was systematic rotations represented by both random and fixed unitary operators. These reflect the effects of gate calibration errors and control field fluctuations. The second type of Markovian noise was amplitude damping, which can represent the process of spontaneous emission. The non-Markovian noise we considered was $1/f$ noise and leakage to higher levels outside the qubit manifold. $1/f$ noise is ubiquitous in nature \cite{Johnson,Schottky,Levitin} and is present to some degree in most physical implementations of qubits \cite{Wellstood1987,Yoshihara2006,Slichter2012,Taylor2007,Pla2013}, although its relative importance is implementation dependent. Leakage can plague a variety of systems, including transmons \cite{Koch2007}, phase-qubits \cite{Martinis}, and quantum dots \cite{Medford2013}. We discuss new models of the fidelity decay curves for $1/f$ and leakage noise and perform numerical simulations under these models.

The structure of our paper is as follows. In section II we introduce the RB protocols and in section III we describe the simulation methods used throughout the presentation. Section IV provides a theoretical and numerical analysis of finite sampling effects in RB, and provides results for simulations of standard RB with Markovian noise. In section V we describe the model and simulation of $1/f$ noise, and provide the results of RB for this $1/f$ noise model. In section VI we provide a new theoretical analysis for modeling fidelity decay under a leakage noise model and present numerical results. In section VII we discuss interleaved RB and present results of different noise simulations. We make concluding remarks in Section VIII.


\section{Randomized Benchmarking Protocols}
The underlying idea behind an RB protocol is to apply sequences of randomly chosen gates from some group and measure fidelity decay as a function of sequence length~\cite{Emerson05}. Ideally, this allows one to extract average properties of
the errors associated with implementing these gates in real quantum devices. The standard protocol \cite{Magesan2011,Magesan2012}, which extends \cite{Knill2008}, chooses the gates from the Clifford
group and gives an estimate of the average gate fidelity, $F_{g}$,
or, equivalently, error rate over the group. Interleaved benchmarking \cite{Magesan2012a,Gaebler2012}
extends the standard protocol to estimate the average gate fidelity
of an individual gate.

For some group $\mathsf{G}$ of unitary operations, the general RB protocol is as follows \cite{Magesan2011,Magesan2012}:
\begin{enumerate}
\item Choose gates from the group $\mathsf{G} = \{U_i \}$ to form $K$ sequences of each length
$m$ from some set $\left\{ m\right\} $ of sequence lengths.
\item For each sequence ${U}_{1},\ldots,{U}_{m}$, determine the gate ${U}_{m+1}=\left({U}_{m}\ldots {U}_{1}\right)^{\dagger}$.
\item Apply each sequence  ${U}_{1},\ldots,{U}_{m+1}$ to some initial state
$\rho_{i}$, measure the output state $\rho_{f}$, and repeat several
times to determine the survival probability of some output state for each
sequence.
\item Average this survival probability over all sequences of the same length, and fit to a pre-determined model.
\item From this model determine the desired quantities of the map.
\end{enumerate}

Unless otherwise specified, quantum channels will be expressed in the Pauli Transfer Matrix (PTM) representation \cite{Chow2012}, in which a matrix corresponding to the quantum channel $\mathcal{R}$ on the space of density operators on the $d$-dimensional Hilbert space of an $n$-qubit system is defined such that $\rho'=\sum_{i,j}\mathcal{R}_{ij}P_{i}\mathrm{tr}\left(P_{j}\rho\right)/d$ where $P_{0}=I^{\otimes n}$, $P_{1}=I^{\otimes n-1}\otimes X$, $P_{2}=I^{\otimes n-1}\otimes Y$, etc. For dimension $d=2^n$, the representation is of dimension $d^{2}\times d^{2}$, as the density operators on $n$ qubits are spanned by the $d^{2}$ $n$-qubit Pauli operators $\mathbf{P}=\left\{I,X,Y,Z\right\}^{\otimes n}$.  We will use a calligraphic font to denote a quantum channel (or map) and a standard math font for a operator. 

The channel representing the average sequence of length $m$ can be written as 
\begin{equation}
\mathcal{S}^{\left(m\right)} = \frac{1}{K}\sum_{\textbf{i}}S_\mathbf{\textbf{i}}^{\left(m\right)},
\end{equation} 
where the sum is over the $K$ sequences  $\mathbf{i} =(i_{1},...,i_{m})$ with
\begin{equation}
\mathcal{S}_{\textbf{i}} = \mathcal{E}_{\textbf{i}_{m+1}}\mathcal{U}_{i_{m+1}} \left(\prod_{j=1}^{m}\mathcal{E}_{\textbf{i}_{j}}\mathcal{U}_{i_j}.
\right)
\end{equation} Here $\mathcal{E}_{\textbf{i}_{j}}$ is the noise on gate $\mathcal{U}_{{i}_{j}}$ implemented at time $j$ with history $(i_{1},...,i_{j-1})$.  
Since $\{\mathcal{U}_{i_{j}}\}$ is a group, the sequence can be rewritten as \cite{Magesan2012}
\begin{equation}\label{eq:twirlcomponent}
\mathcal{S}_{\textbf{i}} = \mathcal{E}_{\textbf{i}_{m+1}} \left(\prod_{j=1}^{m}\tilde{\mathcal{U}}^\mathrm{T}_{i_{j}}~ \mathcal{E}_{\textbf{i}_{j}} ~\tilde{\mathcal{
U}}_{i_{j}}\right),
\end{equation}  where $\tilde{\mathcal{U}}_{i_{j}}$ is another element of $\mathsf{G}$ and all sequences $\tilde{\mathcal{U}}_{1},\ldots,\tilde{\mathcal{U}}_{m}$ are uniformly distributed within the ensemble.  In the limit where $\mathcal{E}_{\textbf{i}_{j}}$ can be approximated by the average error $\bar{\mathcal{E}}$, the average sequence can be represented by \cite{Magesan2012}   
\begin{equation}
\mathcal{S}^{(m)}  =\bar{\mathcal{E}}~ (\bar{\mathcal{E}}_{\mathsf{G}})^{ m}.
\end{equation} 
Here $\bar{\mathcal{E}}_{\mathsf{G}}$ represents the twirl over the group $\mathsf{G}$ and is given by 
\begin{equation}\label{eq:twirl}
\bar{\mathcal{E}}_{\mathsf{G}} = \frac{1}{|\mathsf{G}|}\sum_{\mathcal{U}\in\mathsf{G}} \mathcal{U}^\mathrm{T}  ~\bar{\mathcal{E}} ~ \mathcal{
U},
\end{equation}  which is just a group average.  Depending on the group $\mathsf{G}$, this channel can have a simple structure with a small number of parameters which may be determined by fitting the measured fidelities to the fidelity decay model (FDM)
\begin{equation}\label{eq:fidelity}
F(m)=\mathrm{tr}\left[E\mathcal{S}^{(m)}\rho\right]= \tilde{\mathbf{e}}^{\mathrm{T}}  (\bar{\mathcal{E}}_{\mathsf{G}})^m \mathbf{p},
\end{equation} 
 where  $\rho = \mathbf{p}^{\mathrm{T}} \mathbf{P}/d$ represents the initial state and $\tilde E= \tilde{\mathbf{e}}^{\mathrm{T}}\mathbf{P}$ represents the measurement ($E$)  and the error in the final gate ($\bar{\mathcal{E}}$).
 
From the above there are two important assumptions that need to be addressed.\\
\\
\textbf{Assumption 1: Finite Sampling.} The sample average fidelity converges to the average over all possible sequences for small sample sizes. Because of the length of the sequences used, it is infeasible to implement more than a very small fraction of all possible sequences of each length.\\
\\
\textbf{Assumption 2: Noise Homogeneity.} The average variation of the errors is weak so that most errors are close to the average. In practice, this may not always be satisfied since the errors may have strong gate-dependence (calibration errors, etc.), time-dependence (control field power fluctuations, etc.), or history-dependence (leakage to higher levels, 1/f noise etc.).

\subsection*{Standard Clifford Benchmarking}
Standard Clifford randomized benchmarking (SRB) estimates the average error rate of the errors on the members of the full n-qubit Clifford group. The gates are chosen from this group and Schur's Lemma tells us that Eq. (\ref{eq:twirl}) gives the depolarizing channel
\begin{equation}
\bar{\mathcal{E}}_{\mathsf{G}}=\begin{pmatrix}
1\\
 & \alpha\\
 &  & \ddots\\
 &  &  & \alpha
\end{pmatrix},
\end{equation}
where the basis is ordered such that $I$ is first. The system is prepared in any initial state and the FDM, Eq. (\ref{eq:fidelity}), becomes
\begin{equation} \label{eq.FDM}
F=A\alpha^{m}+\tilde e_0
\end{equation}
where the constants $A  = \sum_{j\neq0} \tilde e_j p_j$ and $\tilde e_0$ absorb
all SPAM errors. In the case that there are no SPAM errors $\tilde e_0=1/d$ and $A=1/d$.    As shown in Ref. \cite{Magesan2012} the average error rate is estimated by $\hat r=\left(1-\alpha\right)\left(d-1\right)/d$.

\subsection*{Interleaved Randomized Benchmarking}
Interleaved randomized benchmarking (IRB) allows estimation of the error on an individual gate, ${U}_{\mathrm{int}}$. The essential idea is to perform two benchmarking experiments; one identical to the standard method described above (which gives the average error depolarizing parameter, $\alpha$, for the Clifford gates $\{\mathcal{U}_{i}\}$), and one in which the gate of interest is inserted (interleaved) between each randomly chosen gate in each sequence to give the depolarizing parameter, $\bar\alpha_{{\mathrm{int}}}$, for the gates $\{\mathcal{U}_{\mathrm{int}}~ \mathcal{U}_{i}\}$.\\
\\
Step 1. Perform Clifford benchmarking as described in the previous section to obtain the average depolarizing parameter $\alpha$ of the errors on the Clifford gates $\mathcal{U}_{i}$.\\
\\
Step 2.  Repeat Step 1, but insert the gate of interest after each of the randomly selected Clifford gates. Then the sequences may be expressed as 
\begin{equation}
\mathcal{S}_{\textbf{i}}=\mathcal{E}_{\mathbf{i}_{m+1}} \mathcal{U}_{i_{m+1}}~\left(\prod_{j=1}^{m}\mathcal{U}_{\mathrm{int}}~\mathcal{E}_{\mathrm{int},j}~\mathcal{E}_{\mathbf{i}_{j}} \mathcal{U}_{i_{j}}\right)
\end{equation} for $\mathcal{E}_{\mathrm{int},j}$ the error on the interleaved gate at time $j$. As before, the group structure permits the sequences to be rewritten as $\mathcal{S}_{\textbf{i}}=\mathcal{E}_{\mathbf{i}_{m+1}}\left(\prod_{j=1}^{m}\tilde{\mathcal{U}}_{i_{j}}^{\mathrm{T}}~\mathcal{E}_{\mathrm{int},j}~\mathcal{E}_{\mathbf{i}_{j}}  \tilde{\mathcal{U}}_{i_{j}}\right)$, which has the same form as Eq. (\ref{eq:twirlcomponent}). The interleave estimate for the depolarizing parameter corresponding to the error on the gate of interest is $\alpha_{\mathrm{int}}=\bar\alpha_{{\mathrm{int}}}/\alpha$. The estimated error rate is $\hat {r}_{\mathrm{int}}=\left(1-\alpha_{\mathrm{int}}\right)\left(d-1\right)/d$ \cite{Magesan2012a,Gaebler2012}. Note that this estimate is provably valid under the following assumption.\\
\\
\textbf{Assumption 3: Product Twirl.} On average, the twirl of the product of two channels is well approximated by the product of twirls. This approximation is exactly correct in the case that at least one of the factor gates is depolarizing, but not in general. A pathological case is when the error on the interleaved gate partially inverts the error on the prior Clifford, in which case IRB can underestimate the error rate. \\
\\
We note that, even when Assumption 3 is not satisfied, it is possible to obtain bounds on the gate error of $\mathcal{U}_{\mathrm{int}}$~\cite{Magesan2012a, Kimmel2013}.

\section{Simulation methods}

In the numerics we only consider single qubit Cliffords, and in this case there are 24 different Clifford operators. A convenient way to decompose these are to introduce the Pauli group, $\mathsf{P} = \{\openone,\mathcal{X},\mathcal{Y},\mathcal{Z}\}$, the exchange group $\mathsf{S}=\{\openone,\mathcal{S},\mathcal{S}^2\}$, and the Hadamard group $\mathsf{H}=\{\openone, \mathcal{H}\}$.  The Pauli group is represented by the maps
 \begin{equation}
 \begin{split}
\mathcal{X}=& \begin{pmatrix} 1 & & &  \\ & 1& &  \\ & &-1 &  \\ & & & -1   \end{pmatrix},
\mathcal{Y}= \begin{pmatrix} 1 & & &  \\ & -1& &  \\ & &1 &  \\ & & & -1   \end{pmatrix},\\
\mathcal{Z}=& \begin{pmatrix} 1 & & &  \\ & -1& &  \\ & &-1 &  \\ & & & 1   \end{pmatrix},
\end{split}
 \end{equation} which just correspond to $\pi$-rotations around the $x$-, $y$-, and $z$-axis respectively. 
The exchange-axis group 
  \begin{equation}
  \begin{split}
 \mathcal{S}= \begin{pmatrix} 1 & & &  \\ & & & 1 \\ & 1 &  &  \\ & & 1&   \end{pmatrix}, 
 \mathcal{S}^2= \begin{pmatrix} 1 & & &  \\ & & 1 &  \\ &  &  & 1 \\ & 1 & &   \end{pmatrix}, 
 \end{split}
  \end{equation} exchanges $(x,y,z)\rightarrow(z,x,y)\rightarrow(y,z,x)$ and the Hadamard group  
\begin{equation}
  \begin{split}
 \mathcal{H}= \begin{pmatrix} 1 & & &  \\ & & & 1 \\ &  & -1 &  \\ & 1 & &   \end{pmatrix}, 
 \end{split}
  \end{equation} exchanges $(x,y,z)\rightarrow (z,-y,x)$. The single qubit Clifford group is the group generated by all combinations of elements in $\mathsf{H}$, $\mathsf{P}$, and $\mathsf{S}$, and has size $2\times 3\times 4 =24$. It is worth noting that the group formed by all combinations of elements in $\mathsf{P}$ and $\mathsf{S}$ is a 2-design consisting of 12 elements, and is the minimum group that can fully depolarize any quantum operation.
  
In many experiments the fundamental operations are $\exp[-iX\theta/2]$ or $\exp[-iY\theta/2]$, which represent rotations around the $X-$ or $Y$- axis by angle $\theta$. In the PTM representation these are represented by  
 \begin{align}
\mathcal{X}_\theta & =\begin{pmatrix} 1 & & &  \\ & 1& &  \\ & &\cos(\theta)& -\sin(\theta)  \\ & & \sin(\theta)& \cos(\theta)  \end{pmatrix}, \\
\mathcal{Y}_\theta & =\begin{pmatrix} 1 & & &  \\ & \cos(\theta)& & \sin(\theta) \\ & &1 &  \\ &-\sin(\theta) & & \cos(\theta)   \end{pmatrix}
\end{align}
 Choosing $\theta = \pi$ gives the Pauli maps $\mathcal{X}$ and $\mathcal{Y}$ respectively, and choosing $\theta = \pm\pi/2$ gives the standard $\mathcal{X}_{\pm \pi/2}$ and $\mathcal{Y}_{\pm \pi/2}$ Clifford generators. Table \ref{tab:cliffords} lists the decompositions of all 24 Clifford elements in terms of both $\mathsf{H}-\mathsf{P}-\mathsf{S}$ and the simple rotations by $\pi$ and $\pi/2$.
 
 \begin{table}
 \begin{ruledtabular}
 \begin{tabular}{|c|l|}
 Clifford's & physical decompostion \\\hline
   $\openone-\openone-\openone$& $\openone$ \\
   $\openone-\openone-\mathcal{S}$& $\mathcal{Y}_{ \pi/2} - \mathcal{X}_{ \pi/2}$\\
   $\openone -\openone-\mathcal{S}^2$& $\mathcal{X}_{ -\pi/2} - \mathcal{Y}_{ -\pi/2}$\\
   $\mathcal{X}-\openone-\openone$& $\mathcal{X}$\\
   $\mathcal{X}-\openone-\mathcal{S}$& $\mathcal{Y}_{ -\pi/2} - \mathcal{X}_{ -\pi/2}$\\
   $\mathcal{X}-\openone-\mathcal{S}^2$& $\mathcal{X}_{ \pi/2} - \mathcal{Y}_{ -\pi/2}$\\
   $\mathcal{Y}-\openone-\openone$& $\mathcal{Y}$\\
   $\mathcal{Y}-\openone-\mathcal{S}$& $\mathcal{Y}_{ -\pi/2} - \mathcal{X}_{ \pi/2}$\\
   $\mathcal{Y}-\openone-\mathcal{S}^2$& $\mathcal{X}_{ \pi/2} - \mathcal{Y}_{ \pi/2}$\\
   $\mathcal{Z}-\openone-\openone$& $\mathcal{X} - \mathcal{Y}$\\
   $\mathcal{Z}-\openone-\mathcal{S}$& $\mathcal{Y}_{\pi/2} - \mathcal{X}_{ -\pi/2}$\\
   $\mathcal{Z}-\openone-\mathcal{S}^2$& $\mathcal{X}_{ -\pi/2} - \mathcal{Y}_{ \pi/2}$\\\hline
   $ \openone - \mathcal{H}-\openone$&$ \mathcal{Y}_{ \pi/2} - \mathcal{X}$\\
    $ \openone - \mathcal{H}-\mathcal{S}$&$ \mathcal{X}_{ -\pi/2}$\\
    $ \openone- \mathcal{H} -\mathcal{S}^2$&$ \mathcal{X}_{ \pi/2} - \mathcal{Y}_{ -\pi/2} -  \mathcal{X}_{ -\pi/2}$\\
    $  \mathcal{X}-\mathcal{H}-\openone$&$ \mathcal{Y}_{ -\pi/2}$\\
    $  \mathcal{X}-\mathcal{H}-\mathcal{S}$&$ \mathcal{X}_{ \pi/2}$\\
    $  \mathcal{X}-\mathcal{H}-\mathcal{S}^2$&$ \mathcal{X}_{ \pi/2} - \mathcal{Y}_{ \pi/2} - \mathcal{X}_{ \pi/2}$\\
    $   \mathcal{Y}-\mathcal{H}-\openone$& $\mathcal{Y}_{ -\pi/2} - \mathcal{X}$\\
    $   \mathcal{Y}-\mathcal{H}-\mathcal{S}$& $\mathcal{X}_{ \pi/2} - \mathcal{Y}$\\
    $  \mathcal{Y}-\mathcal{H}-\mathcal{S}^2$& $\mathcal{X}_{ \pi/2} - \mathcal{Y}_{ -\pi/2} - \mathcal{X}_{ \pi/2}$\\
    $   \mathcal{Z}-\mathcal{H}-\openone$& $\mathcal{Y}_{ \pi/2}$\\
    $   \mathcal{Z}-\mathcal{H}-\mathcal{S}$& $\mathcal{X}_{ -\pi/2} - \mathcal{Y}$\\
    $    \mathcal{Z}-\mathcal{H}-\mathcal{S}^2$& $\mathcal{X}_{ \pi/2} - \mathcal{Y}_{ \pi/2} - \mathcal{X}_{ -\pi/2}$\\
       \end{tabular}
       \end{ruledtabular}
       \caption{A list of the 24 Clifford operators and their decomposition into either physical relevant generators or simple mathematical elements. The operators above the horizontal line form both a group and a 2-design. The $-$ signifies application in time and the mean number of physical generators per Clifford is 1.875 }
       \label{tab:cliffords}
       \end{table}
       
For the benchmarking simulations, unless explicitly noted, the parameter values used were $K=10000$ and $M_{\mathrm{max}}=4096$, with $m\in\{1,2,4,\ldots,M_\mathrm{max}\}$. Exponential fits to the FDM were performed with the MATLAB nlinfit function for the model $A\alpha^m+\tilde e_0$, and $90\%$ confidence intervals were found using the Jacobian option of the MATLAB nlparci function \cite{MATLAB}.

Given a set of RB experiments on a gate set with exact average error rate $r$ and estimated average error rate $\{\hat{r}\}$, we can define the RB accuracy by
\begin{align}
\mu = \log_{10}\left(\hat{r}/r\right),
\end{align}
and the  confidence $C$ by the size of $90\%$ confidence intervals for the fits to the FDM.  The definition of accuracy incorporates logarithms to symmetrically weight multiplicative, rather than additive, deviations of the estimate from the true value.  The average error rate $r$ is defined by 
\begin{equation}
r=1-f_g  =\frac{d^2-\mathrm{Tr}[\bar{\mathcal{E}}]}{d^2+d}.
\end{equation}

\section{SRB: Markovian Errors}

In this section we consider Markovian errors, which correspond to the error map at each time $j$ in the sequence being independent of the previous history of the sequence. These types of errors can arise in a wide variety of scenarios, such as gate mis-calibration (over/under rotation or off-resonant driving), amplitude damping, and control field fluctuations with correlation times much shorter than the individual gate time. 

In order to examine the limits of SRB, we work with quantum error maps (CPTP maps) as far as possible from those on which SRB works best. Up to statistical errors resulting from measurement, SRB exactly estimates the error rate when the error on each gate is depolarizing. We consider the diamond norm distance \cite{Aharonov1998,Kitaev2002,Watrous2005} of an arbitrary gate from a depolarizing channel of the same error rate to be a predictor of benchmarking performance. We used a unitary error model since unitary channels are far from depolarizing with respect to the diamond norm distance (see below).

Random unitary channels of size $N$ were generated by choosing $N\times N$ matrices $S$ and $T$ from the Ginibre ensemble \cite{Ginibre} in which elements are chosen independently from the normal distribution with mean zero and variance one. The unitary channel $U=\exp\left[-iH\epsilon\right]$ is defined with normalized Hermitian matrix $H=\left(G+G^\dagger\right)/\sqrt{\mathrm{tr\left[\left(G+G^\dagger\right)^2\right]}}$ for $G=S+iT$ and where $\epsilon$ is a parameter (found numerically) that gives $U$ the desired error rate $r$. Random CPTP maps acting on density matrices of dimension $d$ were generated by creating a random unitary map of size $d^3$ (with non-normalized Hermitian $H$) and interpreting the matrices $K_{i}\left(j,k\right)=U\left(d^{2}\left(j-1\right)+1,k\right)$, $1\leq i\leq d^{2}$ and $1\leq j,k\leq d$, as a set of Kraus operators.

\begin{figure}[h!]
\centering
\includegraphics[width=.45\textwidth]{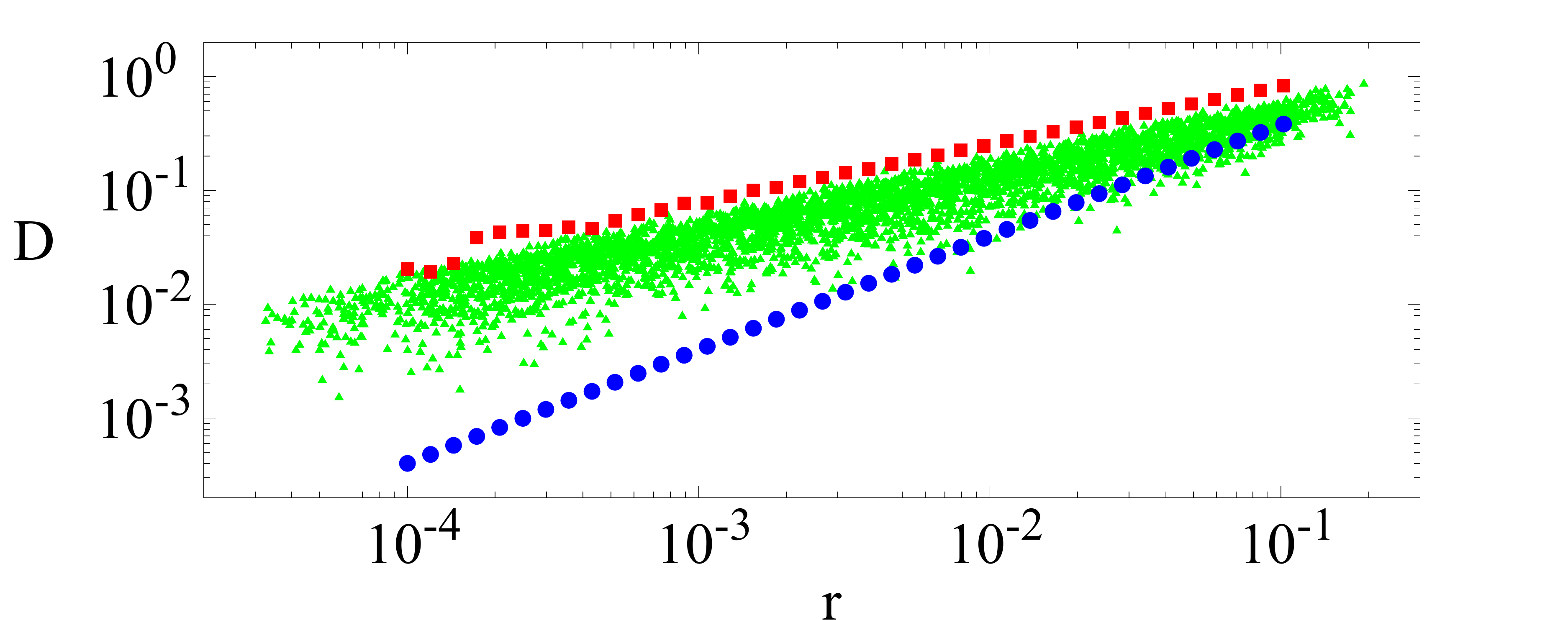}
\caption{(color online) \textbf{Distance of maps from a depolarizing channel}. For a range of error rates, unitary channels (red squares) were farthest from and amplitude damping channels (blue circles) closest to depolarizing channels of the same error rate. Random maps (green triangles) typically fell between these extremes. The distance was measured using the diamond norm distance from a depolarizing channel, and $r$ is the  average error rate. }
\label{fig:DiamondDistances}
\end{figure}

A large number of random unitary and CPTP maps were generated, as were amplitude damping maps of the same error rates. For each map with error rate $r$, the diamond norm distance from the depolarizing channel of error rate $r$ was calculated. As shown in Fig. \ref{fig:DiamondDistances}, we found that unitary channels were farthest from depolarizing channels, whereas amplitude damping channels were particularly close. Therefore, as mentioned above, we consider unitary channels as an adequate worst-case test of SRB.

\subsection{Finite Sampling Effects}

As shown in Ref. \cite{Magesan2012}, the Hoeffding bound can be used to obtain an estimate of the required number $K$ of sequences for a good estimate of the fidelity $F\left(m\right)$ at \emph{each} sequence length $m$. If the trials at each sequence length correspond to i.i.d. random variables with range $\left[a,b\right]$ then

\begin{equation} \label{eq:notrials}
K=\frac{\ln \left({2}/{\delta}\right)(b-a)^2}{2\epsilon ^2}.
\end{equation}
Here $\epsilon$ is the size of the confidence interval and  $1-\delta$ the confidence level. For a $90\%$ confidence level ($\delta = 0.1$) and a confidence interval of $\epsilon=10^{-4}$, we need as many as $K=10^8$ trials for each data point. We will show that this value of $K$ is actually much larger than necessary. The reason for this is that estimation of $\alpha$ (and thus $r$) from a process such as least-squares (LSQ) estimation~\cite{Rencher2012} \emph{simultaneously} uses the information from all data points, whereas the Hoeffding bound analyzes the number of trials for each data point independently. We first provide a simple numerical example from which we see that $K$ can be chosen quite reasonably. Afterwards, we provide a general theoretical result whereby we obtain confidence intervals for $\alpha$ and $r$ using linearization of the non-linear regression model about the LSQ solution. An implication of this result is that $K$ can be chosen to be significantly smaller than the above Hoeffding estimate of $10^8$. 

\begin{figure}[h!]
  \includegraphics[width=0.45\textwidth]{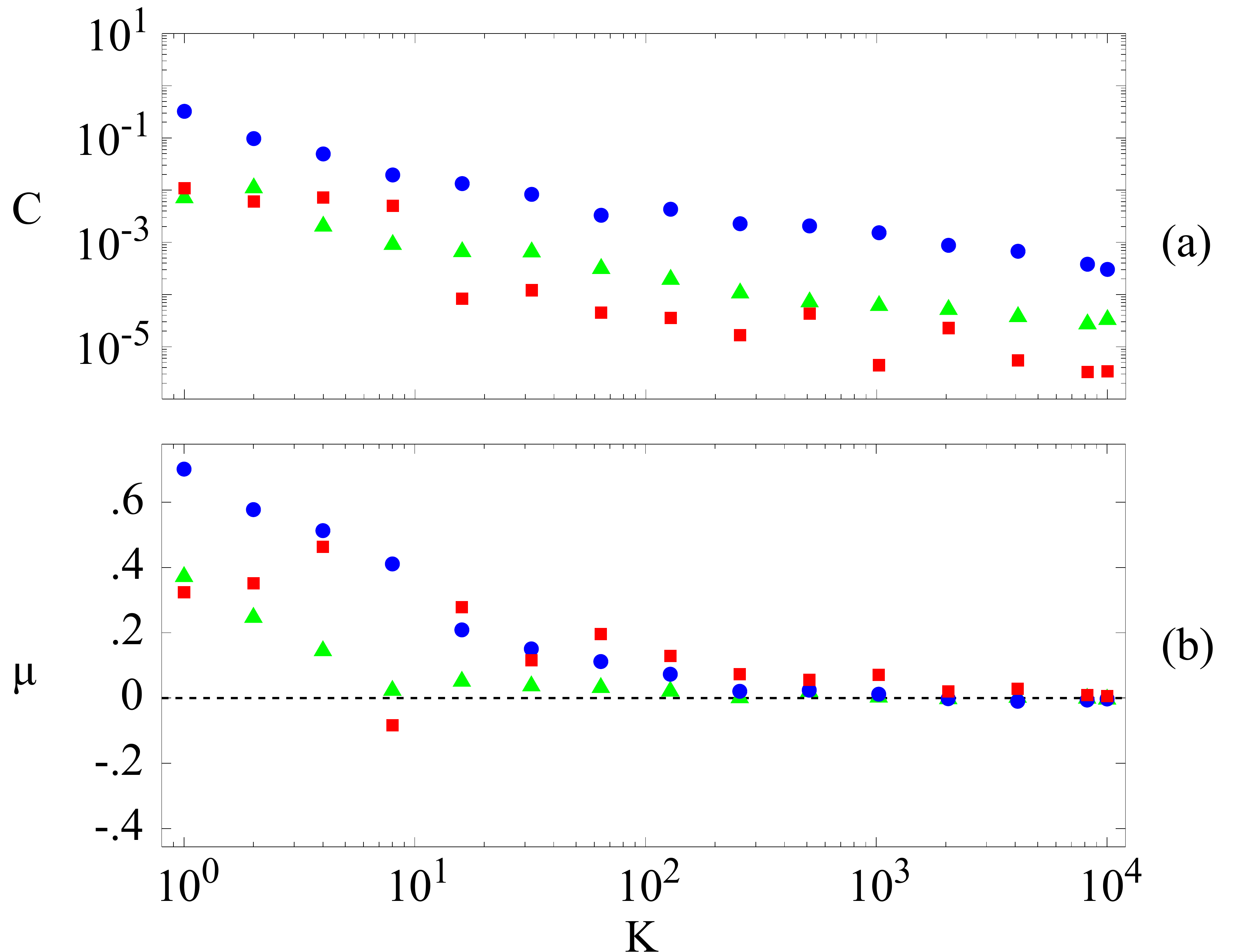}
  \centering
  \caption{(color online) \textbf{SRB with fixed unitary Markovian noise} for error rates $r=10^{-4}$ (red squares), $r=10^{-3}$ (green triangles), and $r=10^{-2}$ (blue circles). (a) Convergence of the confidence $C$ (see text for details). The black line corresponds to the Hoeffding bound method explained in the text. 
  (b) Convergence of the accuracy $\mu$. Note rapid convergence of estimate to within a factor of two of the average error rate by $K=100$.}\label{fig:VaryingKandM}
\end{figure}

For our numerical analysis, we considered various time-independent Markovian error models. Since essentially identical results were obtained for all models, we present the results for the case of gate and time-independent unitary error. Fig. \ref{fig:VaryingKandM}a plots the size of the confidence interval $C$ (at $90\%$ confidence level) on the parameter $\alpha$ for three different error rates: $r=10^{-4}$ (red squares), $r=10^{-3}$ (green triangles), and $r=10^{-2}$ (blue circles). 

From these results, we see that for $K\sim 10-100$, the size of the confidence interval is on the order of the underlying errors, suggesting that fewer than 100 sequences are sufficient to converge to the actual error rate $r$. This is further illustrated in Fig. \ref{fig:VaryingKandM}b) where we plot the accuracy $\mu$ as a function of $K$. This shows that $\hat {r}$ converges (at some $K\sim 10-100$) to within a factor of 2 of $r$. Thus, smaller values of $K$ than $10^8$ suffice for estimating $r$.

We now turn to a more general theoretical analysis based on non-linear regression methods~\cite{Rencher2012}. The FDM Eq. (\ref{eq.FDM}) is a non-linear function with parameters $\{\alpha,A,\tilde e_0 \}$. In the case of linear regression, constructing confidence intervals is exact, however for non-linear regression, confidence intervals are typically constructed via approximative methods. One of the most widely utilized methods, and the approach we take here, is to obtain the least-squares solution, linearize the model around this solution, and construct confidence intervals for the linearization.

In our model there is one predictor variable $m_i$, three parameters we want to estimate $\utheta = (\theta_1,\theta_2,\theta_3) = (\alpha,A,\tilde e_0)$, and one dependent variable $F(m_i,\utheta)$. Let $y_i$ represent the data we acquire so that if $\mathbf{Y}$ represents the vector of $y_i$ values,
\begin{align}
\mathbf{Y}&=\mathbf{F}\left(\uthetat\right) + \tilde{\boldsymbol{\xi}}.
\end{align}
Here, $\tilde{\boldsymbol{\xi}}$ is the realization of the random noise process $\boldsymbol{\xi}$ that produces the observed data, $\uthetat$ is the exact value for the parameters, and $F_i\left(\uthetat\right) = F\left(m_i,\uthetat\right)$. 
We assume for simplicity that each $\xi_i$ is normally distributed with variance ${\sigma^2}/{K}$, where $\sigma$ is the single-shot standard deviation for estimating the fidelity at each sequence length (for simplicity, we assume $\sigma$ is independent of the sequence length).

The LSQ estimator of $\uthetat$ is the vector $\widehat{\utheta}$ that satisfies
\begin{align}
\widehat{\utheta}&=\text{argmin}(S(\utheta))
\end{align}
where
\begin{align}
S(\utheta)&=\left(\boldsymbol{Y}-\boldsymbol{F}(\boldsymbol{\theta})\right)^T\left(\boldsymbol{Y}-\boldsymbol{F}(\boldsymbol{\theta})\right)
\end{align}
Assuming the model in Eq.~(\ref{eq.FDM}) is an accurate description of the fidelity decay, a linearization of $F(m_i,\widehat{\utheta})$ around the LSQ solution $\widehat{\utheta}$ produces a linear model from which we can obtain confidence intervals.

The covariance matrix resulting from the linearized model is given by
\begin{align}
\widehat{V}=s^2\left(J(\widehat{\utheta})^TJ(\widehat{\utheta})\right)^{-1},
\end{align}
where
\begin{align}
s^2&=\frac{S(\widehat{\utheta})}{N-D} \sim \frac{N\sigma^2}{(N-D)K}.
\end{align}
is the average estimated residual variance and $J(\widehat{\utheta})$ is the Jacobian of $\mathbf{F}(\utheta)$ at $\widehat{\utheta}$ which has entries,
\begin{align}
J_{i,j}(\widehat{\utheta})&=\frac{\partial F (m_i;\utheta)}{\partial \theta_j} \Bigg|_{\widehat{\utheta}}.
\end{align}

This gives that, with probability $1-\delta$,
\begin{align}
\widetilde{\theta}_j \in \left[\widehat{\theta_j}-{\widehat{V}_{j,j}}^{\frac{1}{2}}t_{N-D,1-\frac{\delta}{2}}, \widehat{\theta_j}+{\widehat{V}_{j,j}}^{\frac{1}{2}}t_{N-D,1-\frac{\delta}{2}}\right],
\end{align}
where $t_{N-D,1-\frac{\delta}{2}}$ is the $1-\frac{\delta}{2}$'th quantile of the student's t-distribution with $N-D$ degrees of freedom.
$R:=J(\widehat{\utheta})^T J(\widehat{\utheta})$ is a $3\times 3$ matrix
and it is straightforward to calculate 
\begin{align}
Q&=R^{-1},
\end{align}
where we note that $Q$ depends on the fixed parameters $N$ and $\{m_i\}_{i=1}^N$, and on the estimators $\hat{A}$, $\hat{ {e}}_0$, and $\hat{\alpha}$. Since we are mainly interested in estimating $\alpha$, let us focus on
\begin{align}
Q_{1,1} &:= \left(\left[J(\widehat{\utheta})^T J(\widehat{\utheta})\right]^{-1}\right)_{1,1}.
\end{align}
We have,
\begin{align}
\left(\widehat{V}_{1,1}\right)^{\frac{1}{2}}&=s\sqrt{Q_{1,1}}= \frac{\sigma\sqrt{NQ_{1,1}}}{\sqrt{(N-D)K}},
\end{align}
and so, since $D=3$,
\begin{align}
\widetilde{\alpha} \in \left[\hat{\alpha}- \frac{t_{N-3,1-\frac{\delta}{2}}\sigma\sqrt{NQ_{1,1}}}{\sqrt{(N-3)K}}, \hat{\alpha}+ \frac{t_{N-3,1-\frac{\delta}{2}}\sigma\sqrt{NQ_{1,1}}}{\sqrt{(N-3)K}}\right].
\end{align} That is, the confidence interval depends on the standard error $\sigma/\sqrt{K}$ of the experiment. 

Let us now choose a set of parameters that could represent a possible randomized benchmarking experiment. First, suppose we want a 90$\%$ confidence so that $\delta=0.1$. As well, suppose $N=7$, the set of $m$ correspond to $\{m_i\}_{i=1}^7 = \{2^i\}_{i=1}^7$, and, to calculate $Q_{1,1}$, we take $\uthetahat = \left(0.993,\frac{1}{2},\frac{1}{2}\right)$. This gives $t_{0.05,11}\sim 2.132$ and $\sqrt{Q_{1,1}}\sim 0.0476$ so

%

\begin{align}
\widetilde{\alpha} \in \left[\hat{\alpha} - 0.134\frac{\sigma}{\sqrt{K}}, \hat{\alpha}+ 0.134\frac{\sigma}{\sqrt{K}}\right].
\end{align}
We can now see how different values for $\sigma$ and $K$ affect the confidence interval. Taking $\sigma=0.004$ and $K=50$ implies, with confidence $90\%$,
\begin{align}
\widetilde{\alpha} \in \left[\hat{\alpha} - 7.59\times 10^{-5}, \hat{\alpha}+7.59\times 10^{-5}\right].
\end{align}
Hence, we can see that small values of $K$ (much smaller than those dictated Hoeffding bounds for each data point) will still lead to robust estimates of the error-rate $\hat r$.

\subsection{Results and Discussions}

Here, we consider the performance of SRB with respect to various Markovian noise models. The models that we consider are 
\\\\
\emph{Gate-dependent random unitaries}: A different random unitary error is applied to each Clifford gate. 
\\\\ \emph{Fixed random unitary}: The same random unitary error is applied to all Clifford gates.
\\\\ \emph{Generator-dependent unitaries}: Each Clifford gate was decomposed into a minimum number of generators $X_{\pm \pi/2}$, $Y_{\pm \pi/2}$, $X$, and $Y$ (Tab. \ref{tab:cliffords}) which were each assigned a random unitary error of strength $r/1.875$. Error maps were determined from the decompositions. Note the decomposition is not unique but results don't depend on this choice.
\\\\ \emph{Amplitude damping}: We numerically determined the noisy $\mathcal{X}_{\pm \pi/2}$, $\mathcal{Y}_{\pm \pi/2}$, $\mathcal{X}$, and $\mathcal{Y}$ maps and then, using the decomposition given in Tab. \ref{tab:cliffords}, determined from these error maps the noise on the full Clifford group. The generator gates are usually performed by Rabi rotation around the $X$ or $Y$ axis at rate $\Omega$ for time $t_g = \pi/2\Omega$. Amplitude damping noise of error $r$ is characterized by the rate $\gamma =1/T_1 = 4\Omega ~\mathrm{ln}\left[(1+\sqrt{4-6r})/(3-6r)\right]/\pi$ and the noisy generator maps are given by 
\begin{equation}
\mathcal{X}_{{\pm \pi/2}} = \prod_{l=1}^N\begin{pmatrix}
1 \\
 & \eta\\
 &  & \eta{\cos(\pi/2N)}&\mp \eta{\sin(\pi/2N)})\\
 1-\eta^2 &  &  \pm \eta^2{\sin(\pi/2N)}&  \eta^2{\cos(\pi/2N)}
\end{pmatrix},
\end{equation} 
\begin{equation}
\mathcal{Y}_{\pm \pi/2} = \prod_{l=1}^N\begin{pmatrix}
1 \\
  &  \eta{\cos(\pi/2N)}& &\pm \eta{\sin(\pi/2N)}\\
 & & \eta\\
 1-\eta^2 & \mp \eta^2{\sin(\pi/2N)} &  & \eta^2{\cos(\pi/2N)}
\end{pmatrix},
\end{equation}
where $\eta = (1/(\sqrt{4-6r}-1))^{-1/N}$ (and similarly for $\mathcal{X}$ and $\mathcal{Y}$). $N$ is a numerical parameter which we take to be 2000000.
 \\\\ \emph{Gaussian noise (fast)}: The noise on all gates at time $j$ was $V_j=U^{\epsilon_j}$ for some fixed unitary channel $U$. Time-dependent parameter $\epsilon_j$ was chosen such that $V_j$ has error rate $r_j$ that is normally distributed with mean $r$ and standard deviation $r/4$. 
\\\\  \emph{Slow drift}: This is identical to the fixed unitary case except that the fixed unitary depends on $k$ so that $U_k=U^\epsilon$ for numerically determined $\epsilon$ such that $r_k$ increases linearly from $r/2$ at $k=1$ to $3r/2$ at $k=K$.

\begin{figure}[h]
  \centering
  \includegraphics[width=.5\textwidth]{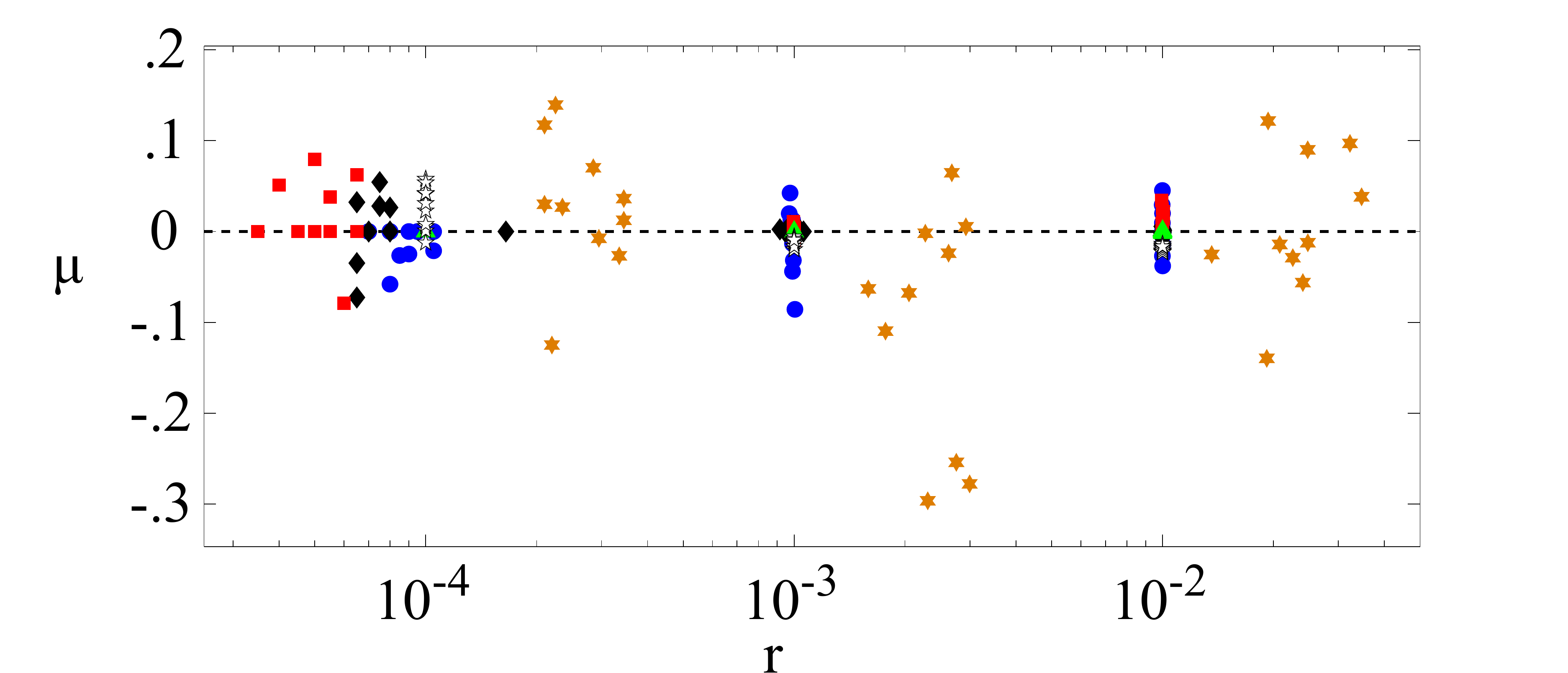}
  \caption{\textbf{The accuracy of several Markovian models} with $K=10000$ sequences.  Different random unitaries (blue circles), fixed random unitary (black diamonds), generator-dependent unitaries (orange 6-pointed stars), amplitude damping (green triangles), Gaussian unitaries (red squares), and Slow Drift (white five-pointed stars). All errors except generator-dependent unitary errors were estimated to within $25\%$ of $r$, with amplitude damping noise determined significantly better. Generator-dependent unitary noise was estimated to within $50\%$ of $r$. The large horizontal spread at low error rates is due simply to the precision of the procedure used for generating random unitary channels of fixed error rate.}\label{fig:Standard Markovian}
\end{figure}

For all noise models tested, SRB estimated the error rate to within a factor of two of the actual average error rate (Fig. \ref{fig:Standard Markovian}, Tab. \ref{tab:AllResults}). SRB performed best with amplitude damping noise, supporting our hypothesis that SRB would work well for errors near the depolarizing channel. Generator-dependent noise was estimated poorest. A possible cause of this, relative to the other Markovian cases, is variation of the error rate over different Clifford gates. This variation may lead to deviations from the exponential FDM, a supposition supported by the larger confidence intervals for this model compared to the others.

\begin{table*}
\begin{ruledtabular}
\begin{tabular}{l|ccc|ccc|ccc}
     {Error rate}
       & \multicolumn{3}{c}{.0001}& \multicolumn{3}{c}{.001} & \multicolumn{3}{c}{.01} \\\hline
       & $\bar\mu$ & $s$ & $\bar C$ & $\bar\mu$ & $s$ & $\bar C$ & $\bar\mu$ & $s$ & $\bar C$\\
      Random unitary & -1.3 $\left[-2\right]$& 5.9 $\left[-3\right]$ & 4.3 $\left[-6\right]$ & -7.3 $\left[-3\right]$& 1.1 $\left[-2\right]$ & 2.2 $\left[-5\right]$ & 2.2 $\left[-3\right]$ & 7.6 $\left[-3\right]$ & 3.9 $\left[-4\right]$\\
      Fixed unitary & 6.6 $\left[-3\right]$ & 1.1 $\left[-2\right]$ & 4.5 $\left[-6\right]$ & 2.2 $\left[-3\right]$& 7.5 $\left[-4\right]$ & 2.3 $\left[-5\right]$ & -1.2 $\left[-3\right]$ & 1.2 $\left[-3\right]$ & 3.5 $\left[-4\right]$\\
      Generator-dependent & 2.7 $\left[-2\right]$ & 2.2 $\left[-2\right]$ & 4.6 $\left[-6\right]$ & -1.0 $\left[-1\right]$ & 3.9 $\left[-2\right]$ & 5.3 $\left[-5\right]$ & 7.0 $\left[-3\right]$ & 2.4 $\left[-2\right]$ & 1.2 $\left[-3\right]$\\
      Gaussian & 1.5 $\left[-2\right]$ & 1.3 $\left[-2\right]$ & 1.2 $\left[-5\right]$ & 1.3 $\left[-3\right]$ & 1.9 $\left[-3\right]$& 2.8 $\left[-5\right]$ & 1.6 $\left[-2\right]$ & 3.0 $\left[-3\right]$& 5.4 $\left[-4\right]$\\
      Slow Drift & 2.5 $\left[-2\right]$ & 7.6 $\left[-3\right]$ & 4.5 $\left[-6\right]$ & -1.2 $\left[-2\right]$ & 1.4 $\left[-3\right]$& 4.0 $\left[-5\right]$ & -1.9 $\left[-2\right]$ &  1.0 $\left[-3\right]$& 4.7 $\left[-4\right]$\\
      Amplitude Damping & 1.7 $\left[-4\right]$ & 9.8 $\left[-5\right]$ & 1.2 $\left[-7\right]$ & 6.9 $\left[-6\right]$ & 3.5 $\left[-5\right]$ & 8.9 $\left[-7\right]$ & -3.8 $\left[-5\right]$ & 1.2 $\left[-4\right]$& 5.2 $\left[-5\right]$\\
      Leakage (random)   & 1.2 [-3]    & 7.4 [-3]    & 3.6 [-6]    & -6.6 [-3] & 5.9 [-3] & 1.4 [-5]  & 6.0 [-4] & 4.5 [-3] & 1.7 [-4]\\
      Leakage (fixed)   &  4.2 [-2] & 2.7 [-2] & 2.8 [-6] & 2.7 [-2]& 1.3 [-2] & 2.0 [-5] & -2.0 [-4] & 1.6 [-2] & 6.0 [-4]\\
      $1/f$   & 7.1 $\left[-2\right]$ & N/A & 7.9 $\left[-6\right]$ & -1.3 $\left[-2\right]$ & N/A &  4.8 $\left[-5\right]$ & -8.4 $\left[-2\right]$ & N/A & 8.5 $\left[-4\right]$\\
      \end{tabular}
      \end{ruledtabular}
      \caption{For each noise model and each average error rate $r$, a set of $n=10$ experiments (with the expection of $1/f$ noise, for which $n=1$), each using $K=10000$ benchmarking sequences was performed. Accuracy $\bar\mu$, standard error $s=(\sqrt{\overline{\mu^2}-\bar\mu^2})/\sqrt{n}$, and average confidence $\bar C$ are as defined in the text.
      The accuracies of the SRB estimates for Markovian errors are in most cases better by roughly an order of magnitude than the accuracies for non-Markovian errors. The precision and fit confidence is not significantly different between the two types of noise, except in the cases of amplitude damping noise, in the presence of which SRB performs especially well. Square brackets indicate multiplication by a power of 10, i.e. $A[x]=A\times 10^x$.
      }
      \label{tab:AllResults}
      \end{table*}

\section{SRB: $1/f$ Noise}

We model a one qubit system subject to semi-classical phase noise by the
Hamiltonian
\begin{equation}
H(t)/\hbar = \frac{\Omega_X(t)}{2}X + \frac{\Omega_Y(t)}{2}Y + \xi(t)Z,
\end{equation}
where $\Omega_{X,Y}(t)$ are real control fields and $\xi(t)$ is a
realization of a real random noise process. The noise process
$\xi(t)$ can be characterized by its power spectral density (PSD)
\begin{equation}
S(f) = \int_{-\infty}^\infty C(t)e^{-i2\pi ft}dt
\end{equation}
where
\begin{equation}
C(t) = \lim_{T\rightarrow\infty}\frac{1}{T}\int_{-T/2}^{T/2}\xi(s)\xi(s+t)ds
\end{equation}
is the autocorrelation function. The noise is said to
be $1/f$ if its PSD is given by $S(f)=A/f$ for some constant $A$.

A simple discrete model of $1/f$ noise is obtained by summing a
large number of random telegraph noise (RTN) realizations with 
different switching rates \cite{Kaulakys}.
A two-state telegraph noise signal $s_k(t)$ switches between $\{+1,-1\}$
with constant rate $f_k$, and inter-arrival times $\tau$ of
switching events are exponentially distributed with probability
distribution $p(\tau)=f_k e^{-f_k\tau}$. If the density of switching
rates is proportional to $1/f$ in the interval
$[f_{\mathrm{min}},f_{\mathrm{max}}]$, then $\xi(t)=A'\sum_k s_k(t)$
has PSD
\begin{equation}
S(f) \propto \frac{1}{\pi f}\left[ \arctan\left(\frac{f_{\mathrm{max}}}{\pi f}\right) - \arctan\left(\frac{f_{\mathrm{min}}}{\pi f}\right) \right],
\end{equation}
which is proportional to $1/f$ if 
$f_{\mathrm{min}}\ll \pi f \ll f_{\mathrm{max}}$ 
\cite{Kuopanportti}.
The noise power is proportional to the square of $A'$
but also depends on the cutoff frequencies and number of RTN signals
participating in the sum.

\subsection{Simulated Ramsey Experiments}

This $1/f$ noise model produces Gaussian decay of coherences \cite{Galperin,Galperin2004,Shnirman}. By simulating Ramsey experiments, we verify
that the model reproduces this type of decay for several values 
of noise power and relate extracted values of $T_2^\ast$ to
average gate fidelities.

Each realization of $1/f$ noise is constructed from $50$ RTN signals
whose initial state is uniformly random. The low and high frequency 
cutoffs are $f_{\mathrm{min}}=(10N\Delta t)^{-1}$ and 
$f_{\mathrm{max}}=(2\Delta t)^{-1}$, respectively, where $\Delta t$ 
is the smallest time step appearing in the simulation and $N$ is
total number of time steps in any noise realization. 

\begin{figure}[h]
  \centering
  \includegraphics[width=0.45\textwidth]{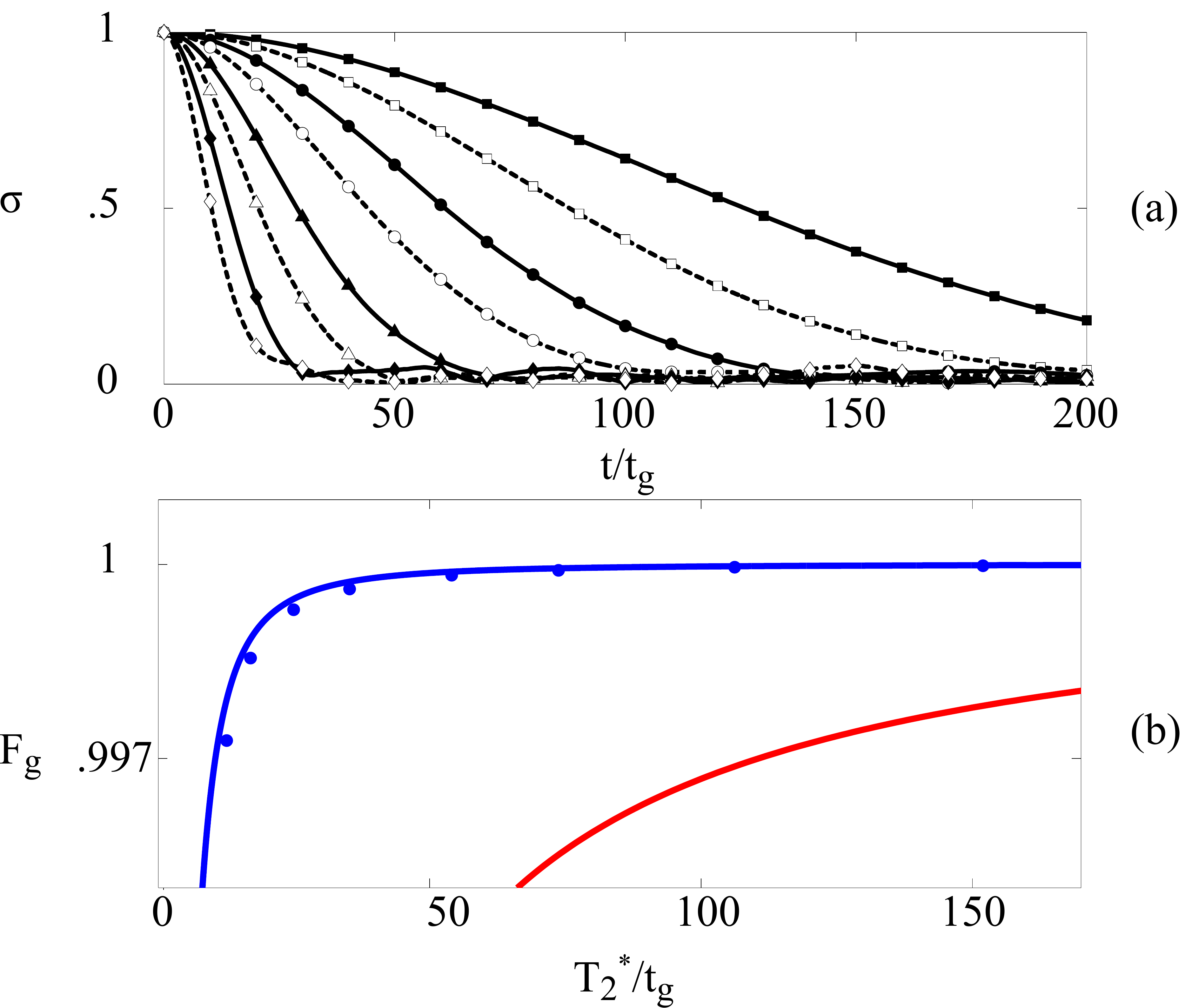}
  \caption{(color online) \textbf{Simulated Ramsey experiments for $1/f$ noise}
  (a) The coherence $\sigma\left(t\right)=2\left|\rho_{12}(t)\right|$ is
  plotted versus time $t/t_g$ where $t_g$ is the duration of
  a gate. Each curve corresponds to a different noise power. The dashed 
  lines interpolate the numerical data and every tenth data point is shown.
  (b) For each decay in (a) we extract $t_g$ as the time
  at which $\sigma=1/e$ and plot the corresponding
  average gate fidelity $F_g=(2+\sigma(t_g))/3$ (blue points). 
  The solid red curve is the gate fidelity for stochastic noise 
  with exponential decay $\sigma(t)=e^{-t/T_2^\ast}$ 
  of coherences. The solid blue curve is the (analytic) gate fidelity for $1/f$ noise: $F_g=\left(2+e^{-\left(t_g/T_2^\ast\right)^2}\right)/3$}\label{fig:OneOverFRamsey}
\end{figure}

In a simulated Ramsey experiment, each $\xi(t)$ yields a pure state 
trajectory $|\psi(t)\rangle=e^{-i2\pi Z\int_0^t\xi(s)ds}|\psi(0)\rangle$ 
where $|\psi(0)\rangle=(|0\rangle+|1\rangle)/\sqrt{2}$. 
Taking the ensemble average over $2000$ noise realizations, we 
obtain the mixed quantum state $\rho(t)$ whose coherence 
$\sigma=2|\rho_{12}(t)|$ exhibits Gaussian decay 
$e^{-(t/T_2^\ast)^2}$, as shown in Fig.~\ref{fig:OneOverFRamsey}a
(see also Fig.~\ref{fig:OneOverF Decay}).
The corresponding (noisy) identity gate fidelity is given by
$F_g=(2+\sigma\left(t_g\right))/3$ where $t_g$ is the gate time, here taken
to be $20\Delta t$.
Provided that $T_2^\ast>t_g$, $1/f$ noise leads to higher gate fidelities than 
stochastic dephasing with the same value of $T_2^\ast$ (see 
Fig.~\ref{fig:OneOverFRamsey}b).

\subsection{Results and Discussion}

To perform a single RB experiment subject to $1/f$ noise, we choose
a sequence of random Clifford gates $C_1$,$C_2$,$\ldots$,$C_{M_{\mathrm{max}}}$.
For each subsequence $C_1,\ldots,C_m$ an inverting Clifford gate
$\Upsilon_m=(C_m\ldots C_1)^\dagger$ is determined.
The subsequences and inverting gates are concatenated to form the 
total sequence
$C_1$,$\Upsilon_1$,$C_1$,$C_2$,$\Upsilon_2$,$\ldots$,
$C_1$,$C_2$,$\ldots$,$C_{M_{\mathrm{max}}}$,$\Upsilon_{M_{\mathrm{max}}}$
for the RB experiment.
Next each Clifford gate is expressed in terms of generators $G_{\pm \pi/2}$ where
$G\in\{X,Y\}$, and each generator $G_{\pm \pi/2}$ is realized by
a normalized Gaussian pulse with amplitude $\pm\pi/2$ on the
corresponding time interval of the control field $\Omega_G(t)$.
The duration of the $1/f$ noise $\xi(t)$ is the same as the duration of
the total sequence, i.e. the noise has the correct correlations over the 
entire duration of the RB experiment.

For each RB experiment, we generate subsequences of lengths 
$m=1,2,\ldots,2^n,\ldots,M_{\mathrm{max}}$ up to $M_{\mathrm{max}}=4096$. 
A time step $\Delta t$ was chosen such that each normalized Gaussian pulse 
was sampled at $20$ equally spaced points. For calculating cutoff 
frequencies, $N$ was taken to be the total number of time steps in the 
experiment. The amplitude $A'$ of the $1/f$ noise was adjusted to 
achieve target average gate error rates $r$ of approximately $10^{-4}$, 
$10^{-3}$, and $10^{-2}$ corresponding to $T_2^\ast/t_g$ values of $94$, 
$30$, and $8$, respectively. Finally the time evolution was calculated 
using the time-ordered composition of discrete unitary gates 
$U(t+\Delta t,t)=\exp\left[-i2\pi H(t)\Delta t/\hbar\right]$.

\begin{figure}[h]
  \centering
  \includegraphics[width=.48\textwidth]{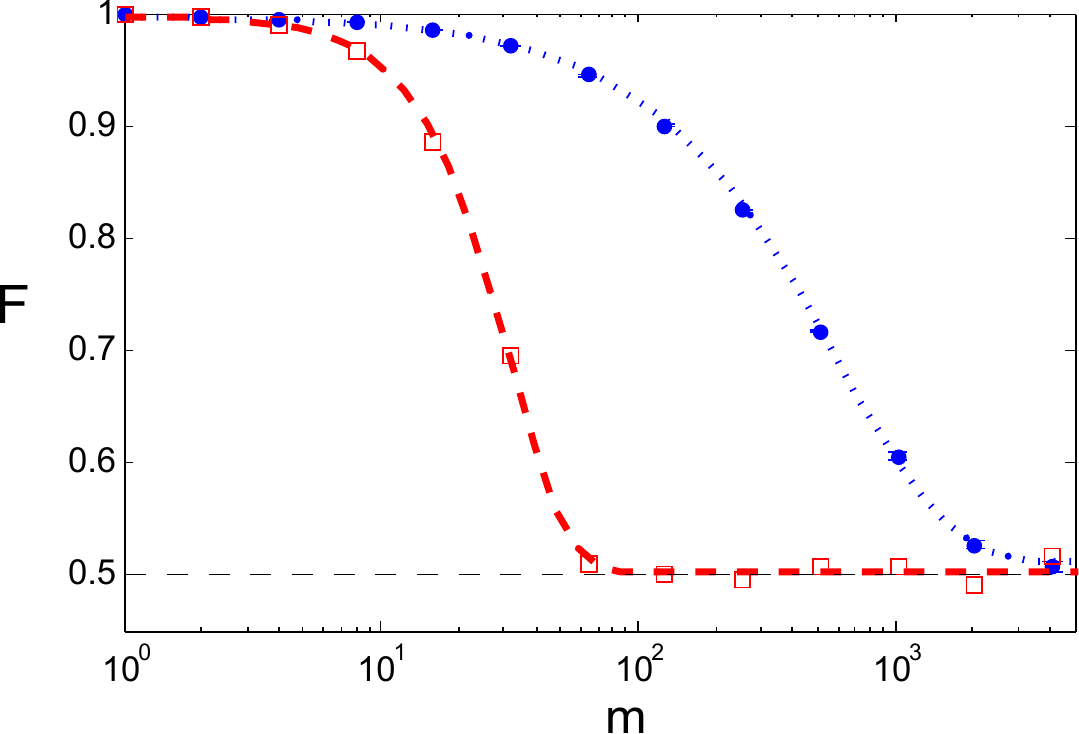}
  \caption{(color online) 
  {\bf Fidelity decay in the presence of $1/f$ noise}
  ($A'=2\times 10^{-8}\rightarrow r\sim 10^{-3}$) with application of
  random Clifford gates (blue circles, mean of four experiments $K=2500$) and identity gates (red squares, $K=2500$). Also shown are fits to the FDM $F=0.486(0.998)^m+0.510$ (blue dotted line) and to the Gaussian model $F=0.495(0.999)^{m^2}+0.503$ (red dashed line). The standard error in the data is less than $4.8\times 10^{-3}$ and fit parameters are rounded to three digits.}
  \label{fig:OneOverF Decay}
\end{figure}

A qubit initialized in the $X-Y$ plane and subjected to $1/f$ phase noise 
suffers a rapid Gaussian decay of state fidelity, but this decay is
dramatically slowed by performing random Clifford gates 
(Fig. \ref{fig:OneOverF Decay}). Insofar as a computation 
may be modeled as a random sequence of Clifford gates, the relevant 
quantity for discussing computational errors may be the gate fidelity 
$F_g$ or, equivalently, the error rate $r$, rather than the dephasing 
time $T_2^\ast$. Supporting this notion is the fact that, 
for a fixed value of $T_2^\ast$, $1/f$ noise has a much higher gate fidelity 
than does noise that exhibits an exponential Ramsey 
decay (Fig. \ref{fig:OneOverFRamsey}).

This behavior is potentially due to the depolarizing effect of
twirling the $1/f$ noise with Clifford gates. Consider a model
with instantaneous Clifford gates followed by noise.
We fix a noise realization $\xi(t)$ and average over SRB sequences.
The noise gives rise to a sequence of correlated random variables
whose samples are $\phi_j=\int_{jt_g}^{(j+1)t_g}\xi(s)ds$,
where $j$ labels the Clifford gate in an SRB experiment.
The noise operators
$e^{-i2\pi\phi_jZ}$ can be twirled independently and become 
depolarizing channels ${\mathcal E}_{\phi_j}$ with depolarizing parameter
\begin{equation}
\alpha_j=(1+2\cos(4\pi\phi_j))/3. 
\end{equation}
Defining $\uphi_m=(\phi_{m-1},\phi_{m-2},\dots,\phi_1)$,
a sequence of $m$ Clifford gates therefore produces the channel
\begin{equation}
\int_{\uphi_m} p(\uphi_m){\mathcal E}_{\phi_{m-1}}{\mathcal E}_{\phi_{m-2}}\dots{\mathcal E}_{\phi_1}{\mathcal E}_{\phi_0}d\uphi_m
\end{equation}
where the joint distribution does not factor into
$p(\phi_{m-1})p(\phi_{m-2})\dots p(\phi_1)$ due to the
low frequency components of the noise.

\begin{figure}[h]
  \centering
  \includegraphics[width=.5\textwidth]{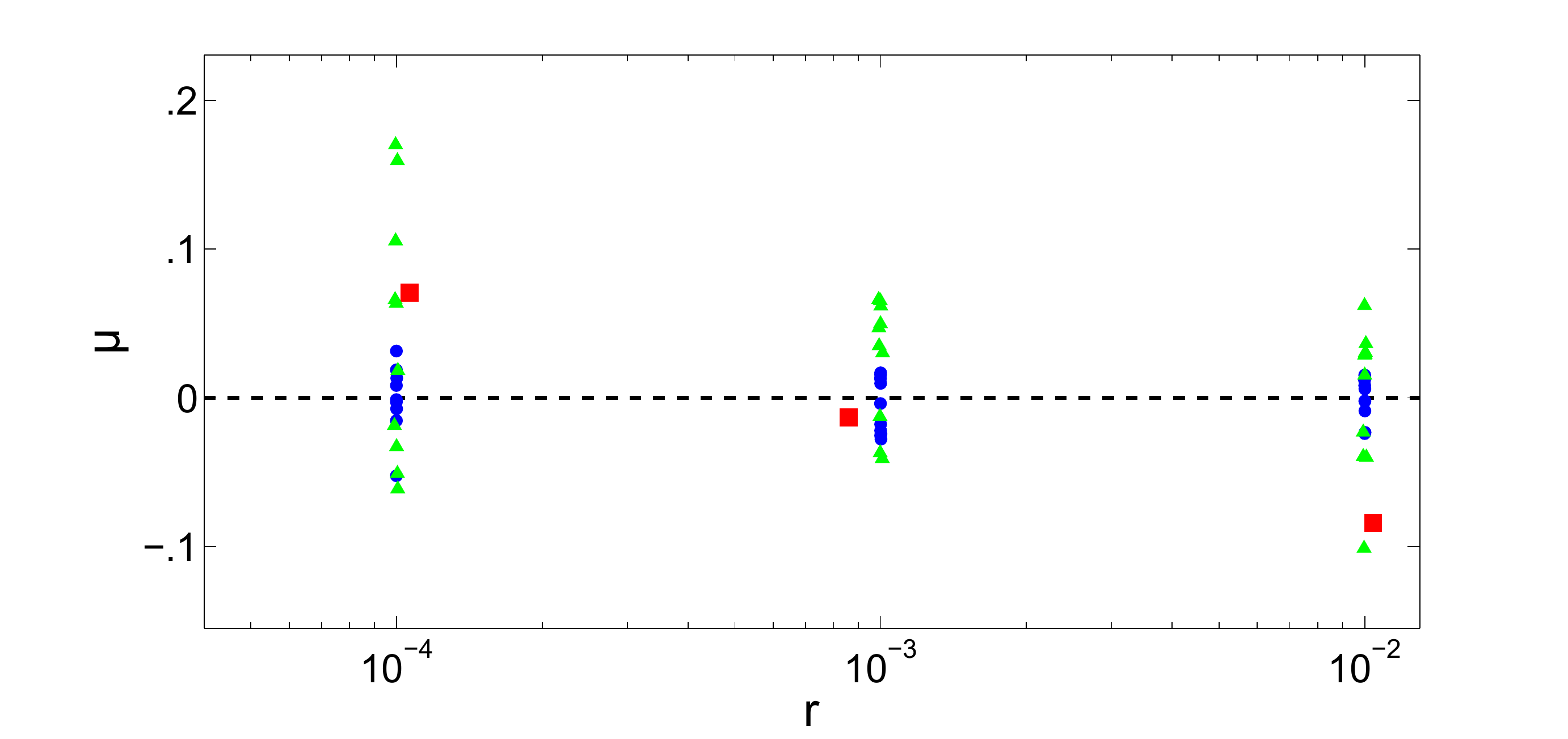}
  \caption{(color online)
  {\bf SRB results for several non-Markovian error models} and error rates
  with $K=10000$ averaged sequences. Leakage, both 
  random and fixed, was nearly always estimated to within a factor of two, 
  and there was no significant difference between the two types, except for 
  the largest error rate, where random leakage had better accuracy, 
  precision, and fit confidence. 
  $1/f$ noise was estimated in all cases to within $25\%$ of $r$.
  Random leakage (blue circles), fixed leakage 
  (green triangles), $1/f$ (red squares).}
  \label{fig:Standard Non-Markovian}
\end{figure}

\begin{figure}[h]
  \includegraphics[width=0.45\textwidth]{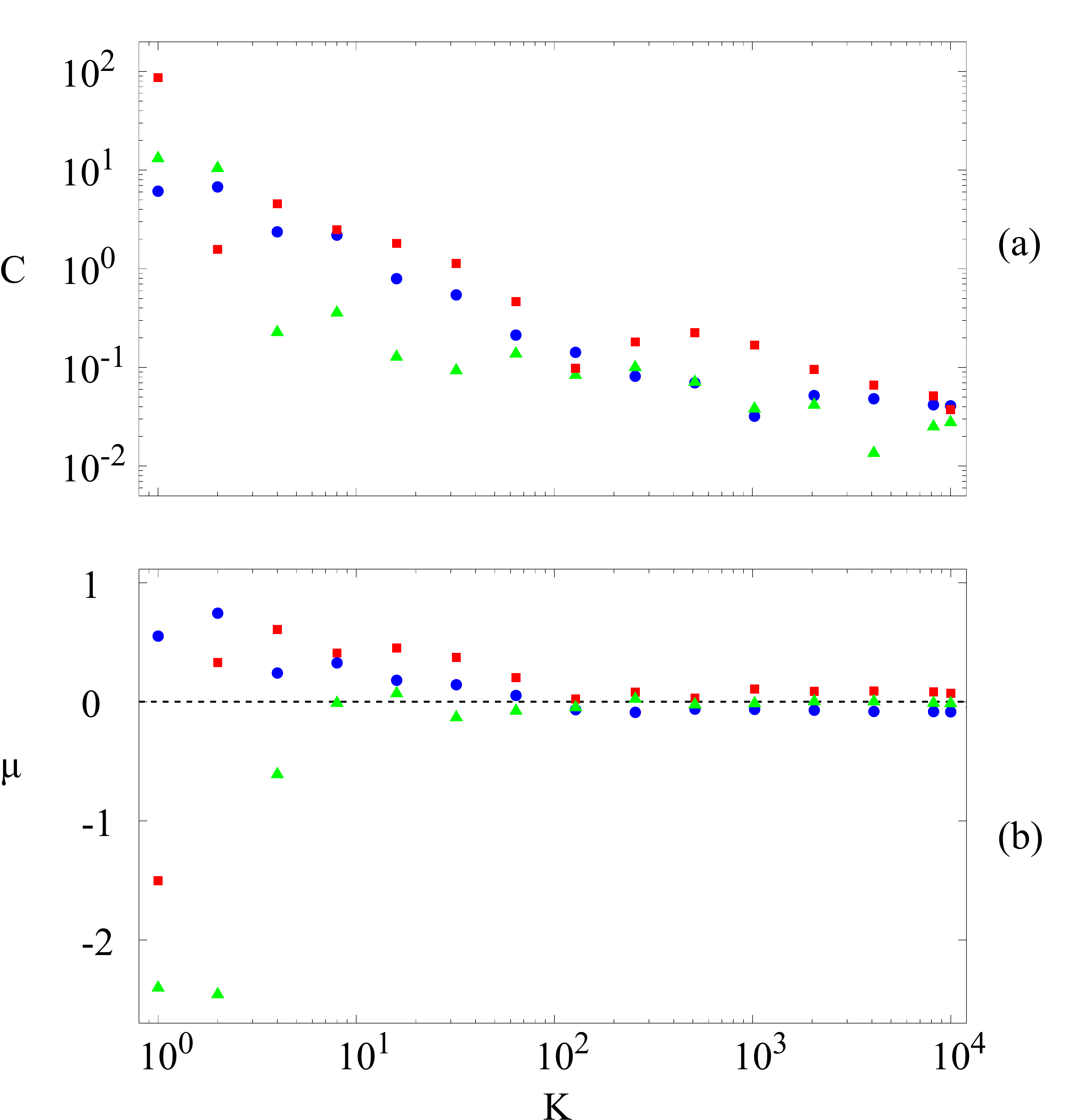}
  \caption{(color online)
  {\bf Convergence in $K$ for standard RB with $1/f$ noise} of
  (a) accuracy $\mu$ (b) confidence interval $C$ on fit of the 
  fidelity decay to the exponential model. Noise strengths are
  $A'=2.5\times 10^{-9}$ $\left(T_2^\ast/t_g=93.95\right)$ [red squares],
  $A'=2.0\times 10^{-8}$ $\left(T_2^\ast/t_g=30.05\right)$ [green triangles],
  and $A'=2.5\times 10^{-7}$ $\left(T_2^\ast/t_g=8.35\right)$ [blue circles].
  }
  \label{fig:OneOverF}
\end{figure}

This behavior foreshadows the result that, for all cases tested, 
RB provides an estimate within a factor of two of the actual average error
rate for $K\sim 100$ or greater (Fig. \ref{fig:Standard Non-Markovian}).
Note, however, that the confidence interval of the RB estimate in the 
presence of $1/f$ noise seems to saturate as $K$ increases, and accuracy 
ceases to improve (Fig. \ref{fig:OneOverF}). This indicates that the 
exponential model of fidelity decay is not completely accurate, in 
contrast to the Markovian case. As a consequence, there is some 
$K_\mathrm{max}$ such that further increases in sample size will not 
yield a more accurate estimate.

\section{SRB: Leakage}

\begin{figure}[h]
  \centering
  \includegraphics[width=.5\textwidth]{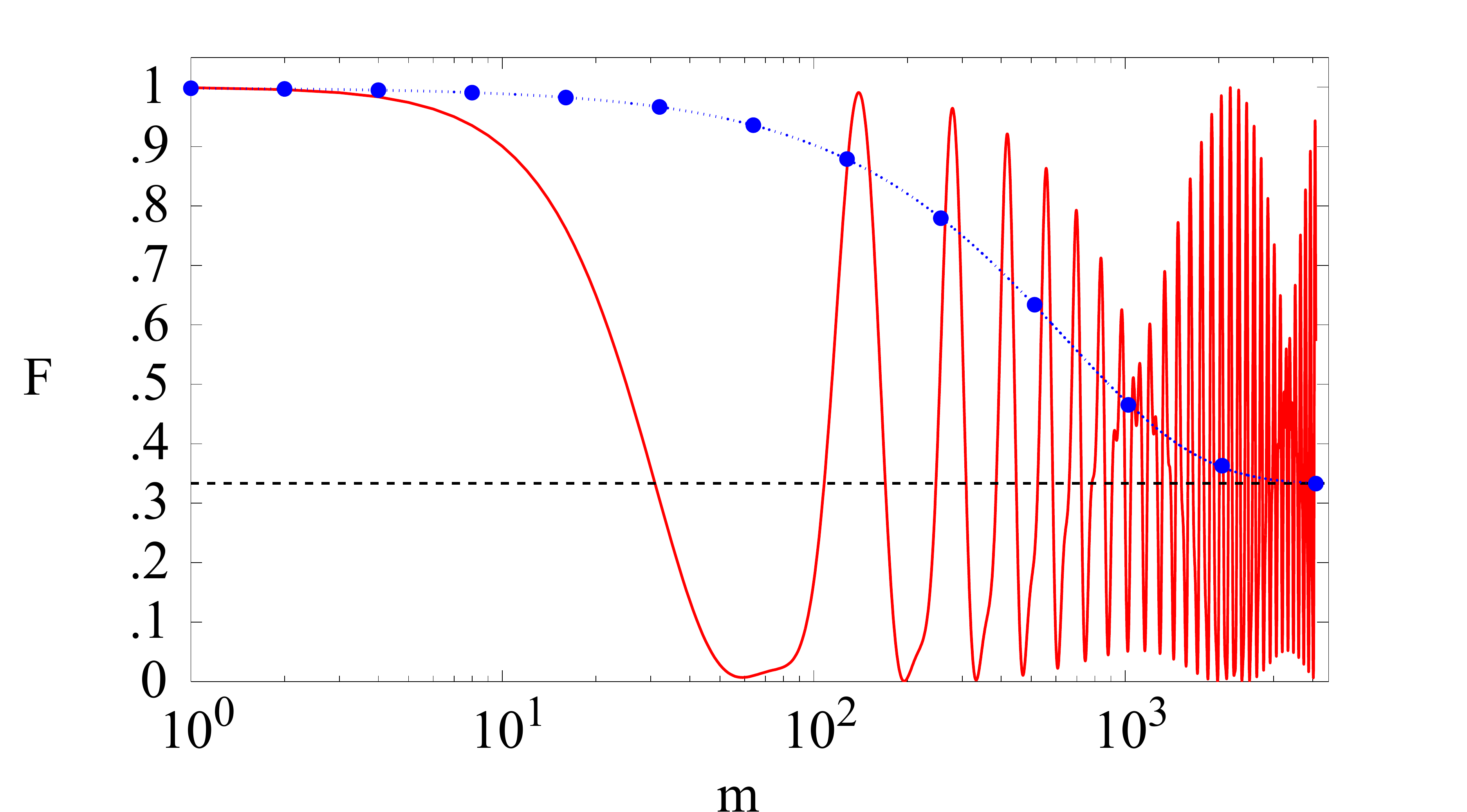}
  \caption{(color online) \textbf{Fidelity decay in the presence of random leakage noise} ($r=10^{-3}$) with application of random Cliffords (blue circles, $K=10000$) and identity gates (red line). Blue dotted line is $F=.666(.998)^m+.333$, the FDM fit to the RB data. Note decay to $1/3$ (black dashed line). Standard errors in the data increase monotonically with $m$ from $10^{-6}$ to approximately $10^{-3}$, and fit parameters are rounded to three digits. The solid red line is the survival probability of the excited state for repeated application of a fixed (randomly chosen) three-level unitary error.
}\label{fig:Leakge Decay}
\end{figure}

In order to simulate leakage, we extend the single qubit Hilbert space to include a third level and model the error by a unitary that acts in the full qutrit Hilbert space. Leakage error is interesting because it can build up coherences in the higher levels so that repeated application may lead to highly non-Markovian dynamics. This is illustrated in Fig. \ref{fig:Leakge Decay} where the fidelity of a state after repeated applications of a typical leakage error is plotted a solid red line. Here we see that the leakage error can lead to many collapse and revival processes in the qubit subspace.

To understand how leakage errors enter into a benchmarking experiment we start by extending the Pauli operators to act on the three-dimensional Hilbert space by defining the following orthogonal basis of nine operators $P_1=I$,
\begin{equation}
\begin{split}
P_2 =\sqrt{3/2} \begin{pmatrix}
X & 0\\
 0 & 0
\end{pmatrix},~
P_3 =& \sqrt{3/2} \begin{pmatrix}
 Y& 0\\
 0 & 0
\end{pmatrix}\\ P_4 =\sqrt{3/2} \begin{pmatrix}
 Z & 0\\
  0 & 0
\end{pmatrix},~P_5 =& \sqrt{1/2} \begin{pmatrix}
1& 0 & 0\\
 0 & 1 & 0\\
 0 & 0& -2
\end{pmatrix} \end{split}
\end{equation}
\begin{equation}
\begin{split}
 P_6 =\sqrt{3/2} \begin{pmatrix}
 0& 0 & 1\\
 0 & 0 & 0\\
 1 & 0 & 0
\end{pmatrix},
P_7 =& \sqrt{3/2} \begin{pmatrix}
 0& 0 & -i\\
 0 & 0 & 0\\
 i & 0 & 0
\end{pmatrix}\\ P_8 =\sqrt{3/2} \begin{pmatrix}
 0& 0 & 0\\
 0 & 0 & 1\\
 0 & 1 & 0
\end{pmatrix},
P_9 =& \sqrt{3/2} \begin{pmatrix}
 0& 0 & 0\\
 0 & 0 & -i\\
 0 & i & 0
\end{pmatrix}
\end{split}
\end{equation} The first four represent population and correlations in the qubit subspace, the fifth represents population inversion in the third level, and the last five represent correlations between the qubit subspace and the third level, and drive leakage. Let us normalize each element so that the set forms an orthonormal basis. Partitioning the basis into spaces $\{P_1,P_2,P_3,P_4,P_5\}$ and $\{P_6,P_7,P_8,P_9\}$, we find that Eq. (\ref{eq:twirl}) becomes
\begin{equation}\label{eq:twirl3}\begin{split}
\bar{\mathcal{E}}_{\mathsf{G}} =& \frac{1}{|\mathsf{G}|}\sum_{\mathcal{U}\in\mathsf{G}} \begin{pmatrix}
 \mathcal{
 U}^\mathrm{T} \oplus 1 & 0\\
  0 & \mathcal{U}_\mathrm{L}^\mathrm{T}
\end{pmatrix}  ~\begin{pmatrix}
\mathcal{A}& \mathcal{B}\\
 \textbf{C} & \mathcal{D}\end{pmatrix} \begin{pmatrix}
 \mathcal{
 U} \oplus 1 & 0\\
  0 & \mathcal{U}_\mathrm{L}
\end{pmatrix}\\& =\frac{1}{|\mathsf{G}|}\sum_{\mathcal{U}\in\mathsf{G}}   ~\begin{pmatrix}
 (\mathcal{
 U}^\mathrm{T} \oplus 1) \mathcal{A}  (\mathcal{
 U} \oplus 1)&  (\mathcal{
  U} \oplus 1)\mathcal{B} \mathcal{U}_\mathrm{L}\\
\mathcal{U}_\mathrm{L}^\mathrm{T} \textbf{C} (\mathcal{
 U} \oplus 1) & \mathcal{U}_\mathrm{L}^\mathrm{T} \mathcal{D} \mathcal{U}_\mathrm{L}
\end{pmatrix},\end{split}
\end{equation} where $\mathcal{ U}$ is the map of the Clifford operator in the qubit subspace and $\mathcal{U}_\mathrm{L}$ is the effect of the unitary in the leakage space. $\mathcal{U}_\mathrm{L}$ does not have a simple form and depends on the relative phase between the qubit subspace map and the higher level. It only maps elements from the leakage subspace to other elements in the leakage subspace. 

 It is simple to show that the first element in Eq. (\ref{eq:twirl3}) becomes   
\begin{equation}
\frac{1}{|\mathsf{G}|}\sum_{\mathcal{U}\in\mathsf{G}} (\mathcal{
 U}^\mathrm{T} \oplus 1) \mathcal{A}  (\mathcal{
 U} \oplus 1)= \begin{pmatrix} 1 & & & & \\ & \alpha& & & \\ & &\alpha & & \\ & & & \alpha & \\\mathcal{A}_{51} & & & & \mathcal{A}_{55} \\ \end{pmatrix},
\end{equation} while the other three do not take such a simple form. However, by allowing the phase in the higher level to be both $U\oplus 1 $ and $U\oplus(- 1) $ for all $U$ in the single qubit Clifford group (extending the size of the single qubit Clifford group to 48 elements) then the set of unitaries becomes $\{(\mathcal{U}\oplus1)(\pm\mathcal{U}_\mathrm{L})\}$.  With this addition, the off-diagonal matrices of Eq. (\ref{eq:twirl3}) become zero, that is, the effective map is block diagonal. This addition is simple to implement experimentally, since it only requires including both $\pm \pi/2$ and $\pm 3\pi/2$ rotations in the generating set. Since the map is now block diagonal, if we prepare and measure in the ground state, the FDM becomes
\begin{align}
F&= \frac{\alpha^m}{2} + \frac{1}{3} + \frac{(\mathcal{A}_{55})^m}{6} + \frac{\mathcal{A}_{51}}{3\sqrt{2}}\sum_{j=0}^{m-1} (\mathcal{A}_{55})^j\nonumber \\
&=  C_1\alpha^m + C_2\mathcal{A}_{55}^m + C_3,
\end{align}
where $C_1= 1/2$, $C_2 =  1/6-D$, $C_3=1/3+D$, and $D=\mathcal{A}_{51}/(3\sqrt{2}(1-\mathcal{A}_{55}))$. This is a simple sum of exponentials and including initial state preparation and measurement errors that only act in the qubit subspace only changes the constants (not the functional form). Furthermore, if the leakage error is from a unital operation (which includes unitary operations), then $\mathcal{A}_{51}=0$. From this model we see that leakage causes the fidelity to asymptotically decay to $C_3$, which is equal to $1/3$ for unital noise. When there is no leakage ($\mathcal{A}_{55}=1$ and $\mathcal{A}_{51}=0$), the decay goes to 1/2 as expected from standard RB. This implies that the asymptotic fidelity value can be used an indicator of the type of noise present in the system. 

\begin{figure}[h]
  \centering
  \includegraphics[width=.5\textwidth]{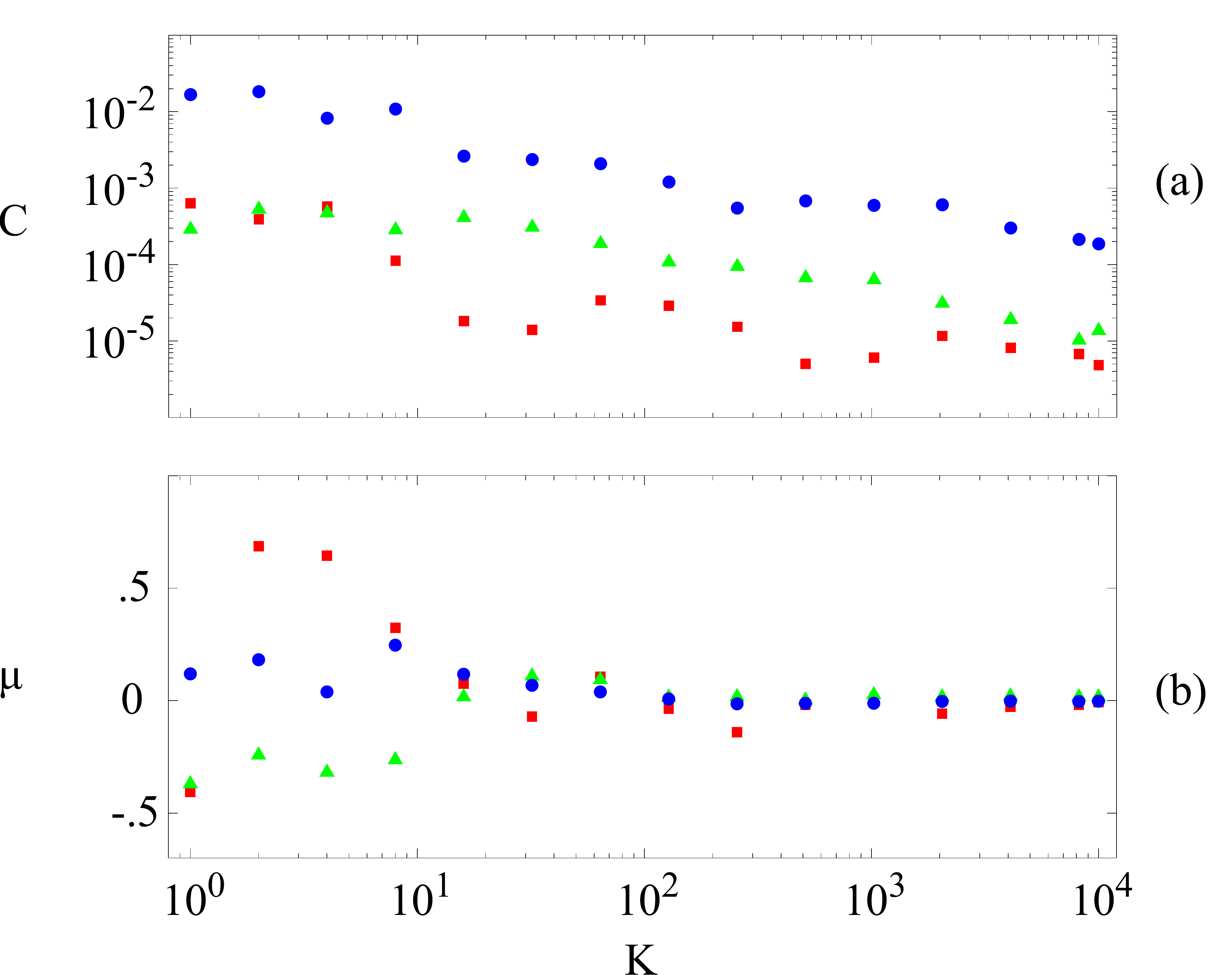}
  \caption{(color online) \textbf{The $90\%$ confidence intervals on the random leakage models} are plotted against sample size $K$ for multiple values of $r=10^{-4}$ (red squares), $r=10^{-3}$ (green triangles), and $r=10^{-2}$ (blue circles).}\label{fig:Leakge Confidence}
\end{figure}

Here we numerically investigate this two-phase model of RB with two types of leakage noise: fixed random and different random three-dimensional unitary channels. A typical benchmarking experiment is shown in Fig. \ref{fig:Leakge Decay} as the blue dotted line for the different random unitary errors. It is clear that the fidelity does not decay to the $1/2$ but rather to $1/3$. For this case the confidence interval and $\mu$ parameter are shown in Fig. \ref{fig:Leakge Confidence}. Here again by about 100 samples the estimated error and the actual error agree. Furthermore, we find that both error models are well captured by the benchmarking experiments and predict the correct underlying error (see Fig. \ref{fig:Standard Non-Markovian}).

\section{Interleaved RB: Markovian Errors}

The goal of IRB is to obtain bounds on the gate fidelity of an individual gate with noise $\mathcal{E}_\mathrm{int}$. In the limit where the average noise $\bar{\mathcal{E}}$ is depolarizing, the bounds collapse and an exact estimate of $\alpha_\mathrm{int}$ is possible. More precisely, for channels $\mathcal{E}_\mathrm{int}$ and $\bar{\mathcal{E}}$ with depolarizing parameters $\alpha_\mathrm{int}$ and $\bar{\alpha}$, if $\bar{\mathcal{E}}$ (or $\mathcal{E}_\mathrm{int}$) is depolarizing, then the depolarizing parameter of the composed channel $\mathcal{E}_\mathrm{int}\bar{\mathcal{E}}$ is given by the product $\alpha_\mathrm{int}\bar{\alpha}$. Typically, one expects that averaging over many sequences implies that $\bar{\mathcal{E}}$ converges to a depolarizing channel and so the depolarizing parameter of $\mathcal{E}_\mathrm{int}\bar{\mathcal{E}}$ converges to $\alpha_\mathrm{int}\bar{\alpha}$.

\begin{figure}[h]
  \centering
  \includegraphics[width=.5\textwidth]{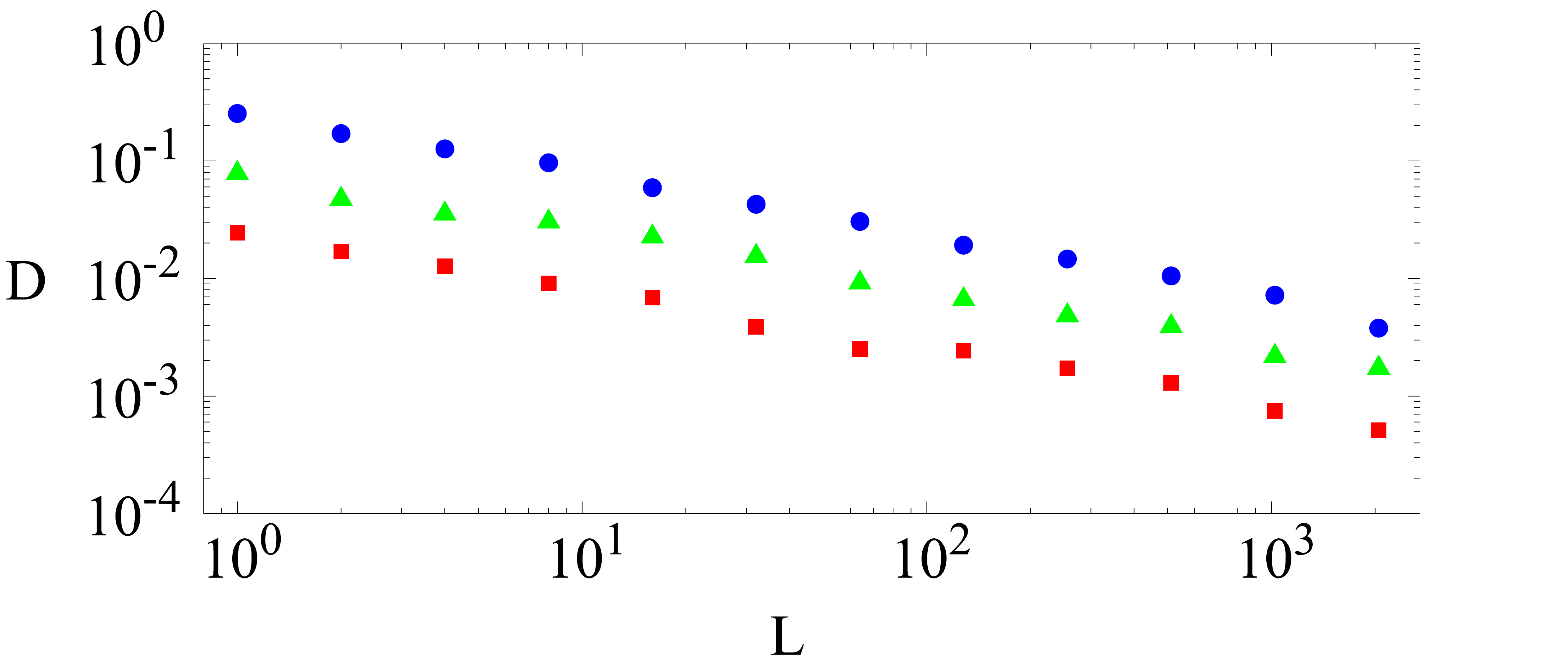}
  \caption{(color online) \textbf{The diamond norm distance $D$ between $\bar{\mathcal{E}}$ and the depolarizing channel} of error rate $r$ decreases as a function of $L$. $r=10^{-4}$ (red squares), $r=10^{-3}$ (green triangles), and $r=10^{-2}$ (blue circles).}\label{fig:AvDiamond}
\end{figure}

We examine the extent to which this is the case by setting $\bar{\mathcal{E}}$ to be an average of $L$ unitary channels. We plot how the diamond norm between $\bar{\mathcal{E}}$ and the depolarizing channel with the same average fidelity scales with $L$ in Fig. \ref{fig:AvDiamond}. We see that the approximation becomes increasingly accurate with increasing $L$. 

\begin{figure}[h]
        \centering
        \includegraphics[width=.5\textwidth]{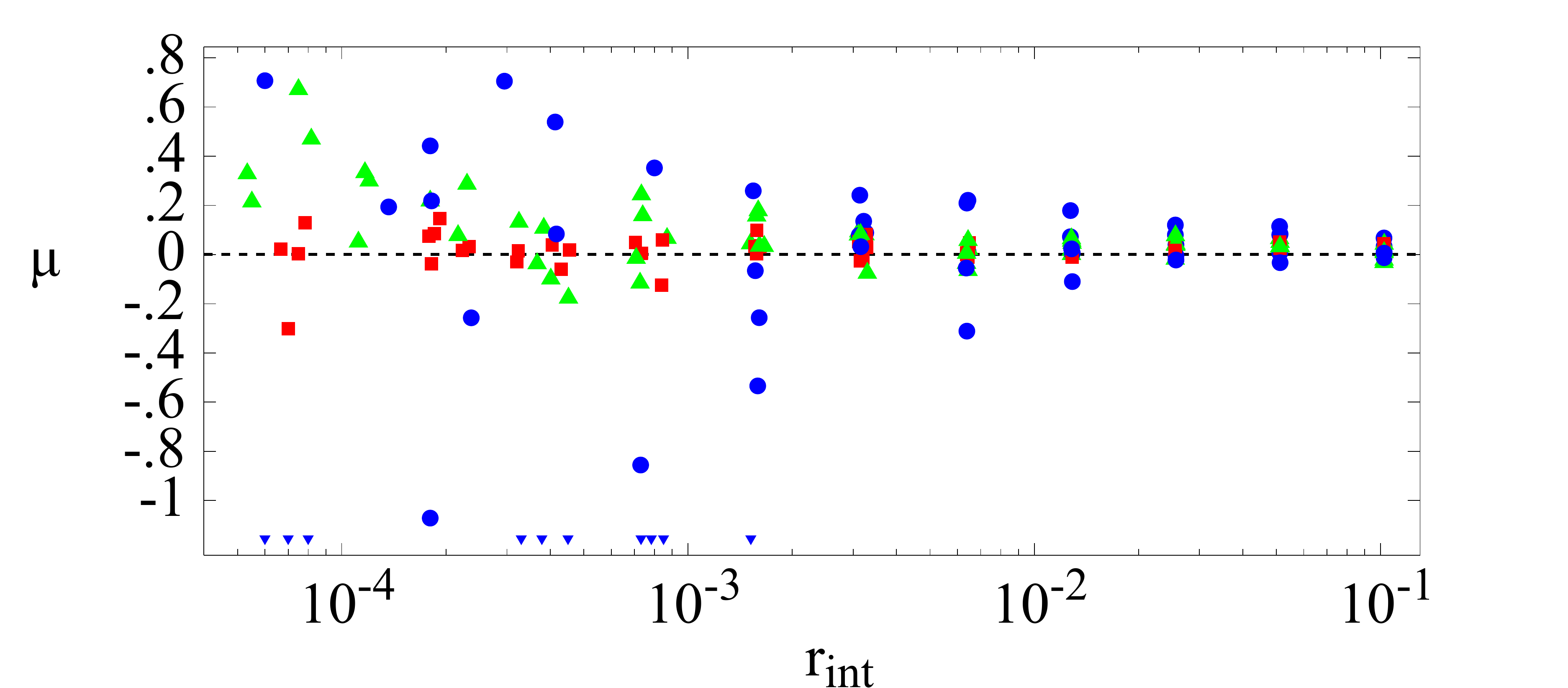}
        \caption{(color online) \textbf{IRB in the presence of random unitary noise for $K=10000$}. $\hat{r}_{\mathrm{int}}$ is the estimated interleaved error rate, $r_\mathrm{int}$ is the true rate, and $\mu=\log_{10}\left(\hat{r}_{\mathrm{int}}/r_\mathrm{int}\right)$. The average Clifford error is: $r=10^{-4}$ (red squares), $r=10^{-3}$ (green triangles), $r=10^{-2}$ (blue circles; blue triangles indicate negative IRB estimates)}.\label{fig:Interleaved}
\end{figure}

IRB was tested using a noise model in which each Clifford gate received a random, gate-dependent unitary error of error rate $r$, and the interleaved gate received a unitary error gate of error rate $r_\mathrm{int}$. The method was tested for three values of $r$ and a wide range of values for $r_\mathrm{int}$ (Fig. \ref{fig:Interleaved}). While the IRB estimates were within a factor of two of the true interleaved error rate for $r_\mathrm{int}\geq r$, they were less accurate as $r_\mathrm{int}$ became much less than $r$ (Fig. \ref{fig:Interleaved}). We can see that when $r_\mathrm{int}/r$ is about $0.1$, the estimate begins to diverge from the true value. Thus, IRB can be a reliable tool in most regimes of interest, unless the interleaved gate is significantly better than a typical gate.

\section{Conclusion}

In this paper we reviewed randomized benchmarking protocols and numerically investigated their application on a single qubit under various physically realistic and relevant error models. These models included systematic rotations, amplitude damping, leakage to higher levels, and 1/f noise. While each randomized benchmarking protocol has a domain of validity for which it provably gives robust error estimates, we found that, in most cases analyzed, benchmarking provides better than a factor-of-two estimate of average error rate. This suggests that RB protocols can be utilized in quite general situations and thus are a valuable tool for verification and validation of quantum operations.

We showed using both numerical and general theoretical results that the number of different random sequences in a benchmarking experiment can be much less than Hoeffding bound estimates~\cite{Magesan2012}. Our theoretical method consisted of finding the non-linear least squares solution, linearizing the non-linear model around this solution, and constructing exact confidence intervals for the linearized multivariate model. We see that the size of the confidence intervals scales linearly with the standard error. 

In the case of $1/f$ noise, we find that randomized benchmarking protocols produce a fidelity decay that can be modeled by a composition of correlated depolarizing channels. The degree of correlation can affect the extent to which a simple exponential decay is valid. For leakage errors, we devised a new protocol that allows for the estimation of gate errors under a sum of exponentials decay model. The asymptotic behavior of fidelity decays can be used as a measure of the extent to which leakage errors are present in an experiment. Finally we showed that, in practice, the interleaved randomized benchmarking protocol provides bounds that are tighter than those theoretically predicted~\cite{Magesan2012a}. Provided the error on the interleaved gate is not much smaller (by a factor of 10) than the average error, the estimated error rate is a reliable quantity.

\begin{acknowledgments}
We acknowledge helpful discussions with Seth T. Merkel, Jerry Chow, and Oliver Dial. We acknowledge support from IARPA under contract W911NF-10-1-0324. All statements of fact, opinion or conclusions contained herein are those of the authors and should not be construed as representing the official views or policies of the U.S. Government.
\end{acknowledgments}


\end{document}